\documentclass[preprint]{aastex}

\received{2001 December 14}
\begin{document}

\title{Far-Infrared and Millimeter Continuum Studies of K Giants: $\alpha$~Boo and $\alpha$~Tau}

\author{Martin Cohen}
\affil{Radio Astronomy Laboratory, 601 Campbell Hall, University of California, Berkeley, CA 94720\\
Electronic mail:  mcohen@astro.berkeley.edu}

\author{Duane F. Carbon}
\affil{NASA/Ames Research Center, Mailstop 258-5, Moffett Field, CA 94035-1000\\
Electronic mail:  dcarbon@nas.nasa.gov}

\author{William J. Welch}
\affil{Radio Astronomy Laboratory, 601 Campbell Hall, University of California, Berkeley, CA 94720\\
Electronic mail:  wwelch@astro.berkeley.edu}

\author{Tanya Lim}
\affil{Space Science Department, Rutherford Appleton Laboratory, Chilton, Didcot OX11 0QX, UK\\
Electronic mail:  T.L.Lim@rl.ac.uk }

\author{Bernhard Schulz}
\affil{IPAC, California Institute of Technology, Pasadena, CA 91125\\
Electronic mail:  bschulz@ipac.caltech.edu}

\author{A. D. McMurry}
\affil{Institute of Theoretical Astrophysics, University of Oslo, Box 1029, Blindern, Oslo, N-0315, Norway\\
Electronic mail: andrew.mcmurry@astro.uio.no}

\author{James R. Forster}
\affil{Hat Creek Observatory, University of California, 42331 Bidwell Road, Hat Creek, CA 96040\\
Electronic mail:  rforster@astro.berkeley.edu}

\author{David Goorvitch}
\affil{NASA/Ames Research Center, Mailstop 245-6, Moffett Field, CA 94035-1000\\
Electronic mail:  David.Goorvitch-1@nasa.gov}

\begin{abstract}
We have imaged two normal, non-coronal, infrared-bright K giants,
$\alpha$~Tau and $\alpha$~Boo, in the 1.4-mm and 2.8-mm continuum
using BIMA.  These stars have been used as important absolute
calibrators for several infrared (IR) satellites.  Our goals are:
(1)~to establish whether these stars radiate as simple photospheres or
possess long-wavelength chromospheres; and (2)~to make a connection
between millimeter-wave and far-infrared (FIR) absolute flux
calibrations.  To accomplish these goals we also present ISO Long
Wavelength Spectrometer (LWS) measurements of both these K giants.
The FIR and millimeter continuum radiation is produced in the vicinity
of the temperature minimum in $\alpha$~Tau and $\alpha$~Boo.  We find
that current photospheric models predict fluxes in reasonable
agreement with those observed for wavelengths which sample the upper
photosphere, namely $\leq$125~$\mu$m in $\alpha$~Tau and $\alpha$~Boo.
We clearly detect chromospheric radiation from both stars by 2.8~mm
(by 1.4~mm in the case of $\alpha$~Boo).  Only additional observations
can determine precisely where beyond 125~$\mu$m the purely radiative
models fail.  Until then, purely radiative models for these stars
can only be used with confidence for calibration purposes below 125~$\mu$m.  
\end{abstract}


\section{Introduction}

We had two distinct objectives in undertaking this work, dominantly to
investigate the merit and reliability of K giant stars as absolute FIR
and millimeter calibration sources, but secondarily to determine just
how shallow in depth simple radiative equilibrium cool star model
atmospheres might be trusted.

We have chosen $\alpha$~Tau and $\alpha$~Boo for this initial study
because these stars, being both bright and nearby, are easily the two
most completely and thoroughly studied red giants.  This makes them
the ideal test cases for our purposes.  It should be noted that they
both lie on the cool, non-coronal side of the ``Linsky-Haisch dividing
line" \citep{linskyhaisch} which separates coronal from non-coronal
giants, although \citet{ayresetal} have found evidence for
buried high temperature emission in the case of $\alpha$~Boo.

\subsection{Infrared calibration stars}
Cohen and colleagues have presented a self-consistent context for absolute
calibration in the IR, ideal for use by ground-based, airborne, and
spaceborne spectrometers and radiometers \citep{i,ii,iii,iv,v,vi,vii,x}.
Their approach is based upon a pair of absolutely calibrated,
IR-customized models of Vega
and Sirius calculated by Kurucz.  These efforts have furnished the
absolute stellar spectrum of Sirius that underpins COBE/DIRBE bands $1-5$
\citep{bcc,ix}, the spectra of K0-M0IIIs for
the on-orbit calibration of the Near- and Mid-Infrared Spectrometers on the
joint NASA/ISAS Infrared Telescope in Space \citep{irts}, likewise
for ESA's Infrared Space Observatory (ISO) \citep{iso}
instruments (e.g., the Short Wavelength Spectrometer) \citep{sws},
for the Spatial InfraRed Imaging Telescope (SPIRIT-III)
on the US Midcourse Space Experiment (MSX) \citep{msx}, and for 2MASS \citep{xiv}.
An extension of this work to substantially fainter stars has provided calibrators 
for the {\it Spitzer} Space Telescope \citep{xiii}.
Most recently, \citet{sdp} have absolutely validated the two calibrated model spectra,
and the empirical spectra of the brighter K giants such as $\alpha$ Tau and $\alpha$ Boo,
in six bands in the mid-infrared (MIR), using precision measurements from the MSX.

The absolute calibration of ISOPHOT \citep{pht,photcal}
was supported by planets, asteroids, and stars, ranked in order of
decreasing flux density \citep{ml,photcal}.  The stellar calibrators
\citep{vii} necessitated long-wavelength extrapolations, as far as 300~$\mu$m, 
of the observed (1.2--35~$\mu$m) absolute stellar spectra by
model atmosphere spectra (computed by one of us
[DFC]: \citet{vii}).  We note that the bright, secondary reference
stars that have been selected for IR calibration purposes are
precisely those ``quiet" giants in which it was suggested that
radiatively-cooled regions very largely dominate the stellar surfaces
so that single component atmospheric models were most likely to be
valid.  Examples of these are $\alpha$~Boo, $\alpha$~Hya, $\alpha$~Tau, 
and $\gamma$~Dra \citep{wied}.  Consequently, one of our objectives
is to establish whether these stars radiate as predicted, thereby
validating the scheme used for ISO at the longest wavelengths, or
possess long-wavelength chromospheres.

\subsection{Linking infrared and millimetric calibration}
The ISO Long Wavelength Spectrometer (LWS) \citep{lws,lwscal}
used Uranus as a calibration standard but also observed Neptune, and
Mars when it became available. Although Uranus has remained the LWS primary
calibrator, the Mars spectrum has been shown to be fully consistent with
this Uranus framework \citep{sidher}. In the case of Neptune, a new
model based on the LWS data has been adopted \citep{orton}. LWS was
able to observe some stellar calibrators \citep{tanya}, although these
observations were difficult to make due to poor signal-to-noise
in the relatively narrow
LWS bandwidths. However, the stellar spectra did demonstrate consistency
with the Uranus-Mars framework, in the LWS 45--170~$\mu$m region. In this paper we
present the totality of results from the ISO LWS spectra obtained on
$\alpha$~Tau and $\alpha$~Boo.

The long-wavelength stellar spectral extrapolations are based upon
the assumption that these calibrator stars are not attended by extensive
chromospheres.  Radiometric testing with ISOPHOT \citep{ml,photcal} 
suggests that these extrapolations were meaningful,
i.e. consistent with calibrations based on asteroids and planets.
It would be of great interest, both scientific and pragmatic,
to determine independently whether these same stellar calibrators radiate as
predicted in the millimeter (mm) domain.
The basis for mm-wave absolute flux calibration is Mars.  Therefore, we
observed the stellar calibrators against Mars, or against
secondary microwave standards that were ultimately traceable to a
comparison with Mars, Uranus, or Neptune.  By this means we hoped to
provide a link between the reference flux density scales of planets and
stars and to unify flux calibration across the IR-mm regime.
Early efforts in this direction at Berkeley were described by \citet{wrixon}.
\citet{gibson4} and \citet{gibson5} have continued this radio absolute calibration work.

\subsection{Cool stellar atmospheres}

Our secondary goal is to probe the structure of the outer atmospheres
in K and, eventually, M giants.  \citet{val76} demonstrated in a
classic paper how observations of the solar FIR continua, atomic
lines, and UV continua could be brought together to construct a
single-component, time-independent mean solar temperature structure.
FIR continuum fluxes were of crucial importance in helping to
establish this mean structure from the upper photosphere, through the
temperature minimum region, and into the chromosphere.  Using the FIR
continuum to probe the shallow depths in solar and cooler stars is
possible because the primary IR continuous opacity, H$^-$ free-free,
increases as $\lambda^2$, pushing the depth of continuum formation
into the outermost stellar layers.  Until now these layers in cool
stars have been accessible only through non-LTE analyses of the cores
of strong atomic and molecular lines (e.g., \citet{al}, \citet{kelch},
\citet{mcmurry}, \citet{wied}).  

While chromospheric models have become more complex in the years since
the \citet{val76} study, the FIR and millimeter fluxes still have the
potential to be valuable diagnostics of the shallow stellar layers.
In accurately determining the FIR and millimeter fluxes of selected
red giants, we hope to provide useful information for testing models
of the cool star stellar photospheres and chromospheres.  We are
particularly interested in determining the location of the transition
from observed fluxes, reproducible by purely radiative equilibrium
models, to fluxes that require invoking a non-radiatively heated
chromosphere.  In addition, we wish to trace the observed run of
decreasing fluxes with increasing wavelength from the upper
photosphere and determine how closely current stellar model atmosphere
calculations can match this run.  This information on the location
of the temperature reversal and the quality of fit to the fluxes
originating in the upper photosphere are vital in defining the
wavelength range for which simple radiative models may be used for FIR
calibrations.

\section{The observations}
Table~\ref{obs} summarizes our various sets of mm observations, listing
array configurations and frequencies used, integration times 
on the stars, phase and flux calibrators
(chosen to be as close to each star's direction in the sky as possible),
and the primary or secondary mm-calibrator observed with
each set of stellar observations.  A few measurements were taken in the
B-array, with which planetary calibrators like Mars were partially resolved,
necessitating an extra step in the calibration process to calculate the flux
expected on the shortest baseline pairs (see below).  Consequently, we
preferred measurements in C- or D-array configurations.  Angular diameters
of these stars in the optical to MIR regions are of order 20~mas,
rendering them unresolved targets in all these configurations, as were the
local calibrators chosen.
On the basis of estimates of the BIMA performance at 1 and 3~mm, our
observing scripts called for typical tracks of several hours' duration,
of which between roughly two and four hours were spent on the stars themselves.


This was a continuum experiment, and we selected observing
frequencies from considerations of the expected stellar spectra,
receiver performance, and atmospheric transmission.  Whenever possible
we included measurements of Mars with each set of stellar and local
calibration source observations, in an attempt to unify the
calibration.  For those few data sets taken in an
array configuration that resolved Mars, we calculated visibilities as a
function of baseline (in kilo$\lambda$)
and applied these partial-resolution factors to the Mars model
\citep{wright} flux densities using our shortest baseline pairs.

Sometimes no planet was visible close in time to our observing
tracks.  Then we used MWC~349 when it was available.
Hat Creek has a long series of observations of this star, it is
relatively bright, has shown no statistically significant variability
in our archives, and has been modeled in detail by \citet{mwc349}
so that flux extrapolation and interpolation to our frequencies is viable.
Further, \citet{gibson5} have obtained its mm flux densities to a
precision of 1\% 
absolute. For our final observations, in December 2000, we used
W3OH as reference for our $\alpha$~Tau track, again because of its relative
proximity to this star, long-term temporal stability of signal, and the
available multifrequency Hat Creek archives which enable interpolation
between known flux densities at other frequencies to our own.  We treated
W3OH as having a canonical H{\sc ii} region spectrum and included the
frequency dependence of the Gaunt factor.  Other secondary references
such as 1415+133 and 0530+135 were likewise observed when no other
obvious flux reference presented itself.  Independent
observations of these were extracted from the Hat Creek archives,
preferably with respect to a primary or acceptable secondary
standard.  In such circumstances, Table~\ref{obs} carries the primary
standard to which the flux pedigree of such a secondary is traceable.

Upper and lower sidebands were processed independently, maps made of
the stars at each frequency, and peak flux densities extracted (see the
final column of Table~\ref{obs}).  Sometimes, weather conditions
were adequate in the lower sideband but led to unacceptably large noise
in the upper sideband.  In such cases we eliminated the noisy datasets
which were readily recognized by the presence of artifacts in the maps, 
such as global striping, or large negative-going blobs. To estimate 
the uncertainties in the tabulated peak flux densities derived from any
map, we used the standard deviation calculated from substantial regions
of the map that excluded the target star itself.
To combine the simultaneous upper and lower sideband maps at one epoch, we
used inverse-variance weighting of the two maps based upon these noise
estimators, relying upon the {\sc miriad} task {\sc maths}.
Likewise, we combined results across epochs.

\section{Self-calibration of the 1-mm data}
Phase calibration of interferometer data relies on transferring phases
accurately from a point source (usually a QSO) to the target source.
The accuracy of the calibration depends on the distance between source
and calibrator, baseline errors, atmospheric phase stability, and the
signal-to-noise ratio (SNR) in the phase measurement of the calibrator.
All sources of phase error result in decorrelation of the
calibrated image; i.e., a lower amplitude in the final map compared with
perfectly calibrated data.  Self-calibration offers a way to eliminate
these phase errors, but is limited by the timescale on which phase can
be accurately measured on the target source itself.

In the case of $\alpha$~Boo, the measured peak flux density in the
normally calibrated image is about 50~mJy.  Given the
system temperature (350~K), bandwidth (670~MHz, reduced from 800~MHz due
to edge-channel flagging in the wideband average), and a correlator
efficiency of 0.88 (2-bit) for these observations, the expected noise in
a single-interferometer measurement is about 70~mJy in 10~min.
Using self-calibration with a 9-element array improves the SNR for
antenna-based phases by about a factor of 3, due to baseline averaging.
The expected r.m.s. noise in the antenna-based phases using
self-calibration is, therefore, about 25~mJy in 10~min.  Assuming the
true flux of $\alpha$~Boo to be
80~mJy, we would expect an SNR of about 3 using 10-min averaging
on the source.  The expected phase error for an SNR of 3 is $\sim20^\circ$
for each antenna (arcsin(1/3)).  The decorrelation expected for
an r.m.s. phase error of 20$^\circ$ is about 6\% 
(reduced amplitude = exp($-r^{2}/2$), where $r$=the r.m.s. phase error 
in radians).  It would, therefore, appear safe to use an
interval of 10 min or longer to self-calibrate the 1-mm $\alpha$~Boo data
in phase.  Of course, any additional phase error (e.g., atmospheric)
incurred during the self-calibration interval will increase the
decorrelation in the final image.

We can estimate the amount of atmospheric decorrelation expected.  The
average r.m.s. delay path in 10~min for a 100~m baseline on the day of
the $\alpha$~Boo 1-mm observations (01 Jun 2000) was 210~$\mu$m, or $\sim$1 rad
at 1360~$\mu$m wavelength.  In D-array, the maximum baseline is about 30~m.
Assuming that the atmospheric r.m.s. phase scales as (baseline)$^{5/6}$,
the r.m.s. phase becomes
0.37~rad on the longest baseline [(30~m/100~m)$^{5/6}$~*~1~rad], and about
0.2~rad for the median baseline of 15~m.  The expected atmospheric
decorrelation in 10~min for the 30~m baseline ($r$=0.37~rad) is 0.85,
and for the 15~m baseline ($r$=0.2~rad) is 0.98.  Thus, atmospheric
decorrelation on 10-min timescales is not expected to be very important,
and phase variations on the longer timescales will be removed by the
self-calibration.

We validated these ideas on the June 2000 observations of $\alpha$~Boo by
experimenting with self-calibration timescales in the reduction of the lower
side-band data.  The total cleaned flux in the resulting images of this star
showed a spread of only 3.5\%, while the more relevant quantity, the stellar
peak flux densities, were 82.1, 81.5, and 81.1 mJy for a 15-, 20-, and 30-min
timescale.  We settled on a 20-min self-calibration timescale to reduce the
data for $\alpha$~Boo, and 30~min for $\alpha$~Tau.

\begin{figure}
\plotone{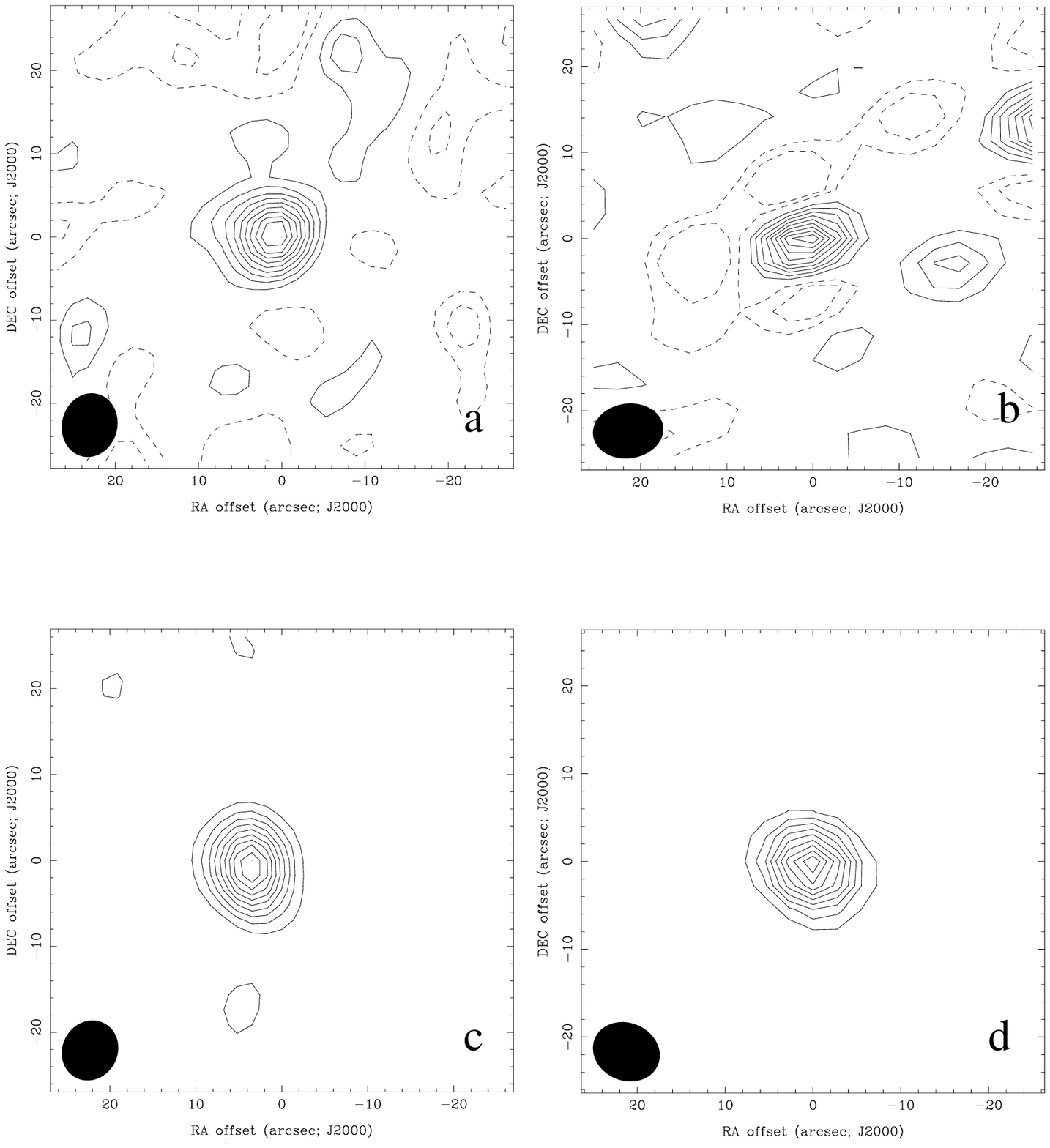}
\caption {Images of both stars at 3~mm and 1~mm constructed from the combination of
all available data. In each plot, contours are given at -20, -10, 10, 20, 30, 40, 
50, 60, 70, 80, 90, 100\% of the peak flux density.
a) $\alpha$~Tau at 3~mm with a 13.97 mJy peak. b) $\alpha$~Tau at 1~mm with a
25.78 mJy peak. c) $\alpha$~Boo at 3~mm with a 20.09 mJy peak. The offset of the 
peak from the stellar position after allowance for proper motion is caused by the 
process of self-calibration. d) $\alpha$~Boo at 1~mm with a 83.5 mJy peak.}
\label{bima13}
\end{figure}

The final rows in Table~\ref{obs} present the combined peak flux density
information based on the totality of the available, good quality, stellar
data for $\alpha$~Tau and $\alpha$~Boo.  The corresponding combined images from
which these final flux densities were determined appear as Figure~\ref{bima13}.

\section{Observations with LWS on ISO}
Observations of $\alpha$~Tau with full wavelength coverage were made in three 
revolutions: 818 (TDT 81803302), 848 (TDT 84801701), and 861 (TDT 86101101). The
observations of $\alpha$~Tau each consisted of 28 scans, 14 in each direction
with a spectral oversampling of 4.  Each observation had about 4630~s integration.
LWS made two AOT observations of $\alpha$~Boo in revolutions 448 (TDT 44800303)
and 608 (TDT 60800904), each with about 2300~s integration.
These observations consisted of 12 scans, 6 in each direction,
taken with an oversampling of 4.

Each LWS spectrum was processed with Offline Processing version 10 (OLP10) and
all the data were reduced using the Infrared Spectral Analysis Package (ISAP).
The reduction steps consisted of removing outliers and averaging over
individual detectors at maximum resolution, clipping at 2.5$\sigma$.
No attempt was made to scale the detectors to make the sub-spectra overlap in the
reductions.  We accomplished this outside ISAP using the splicing techniques
described by \citet{ii} for constructing absolutely calibrated spectra.
Care was taken to use the same number of scans in each direction, so the effects
due to transients in the data are minimal, and no transient correction was
applied.  The fundamental Uranus model for calibrating the LWS is a synthesis of
globally averaged Voyager InfraRed Interferometer Spectrometer (IRIS) data
up to 50~$\mu$m and James Clerk Maxwell Telescope data covering the
0.35--2~mm wavelength range \citep{icarus} with a radiative transfer model
similar to the one described by \citet{icarus}.

A formal off-source, full-grating, LWS integration is available only for $\alpha$~Boo
(TDT 60801005), and two fixed-grating (sampling only ten wavelengths with no detector
scans) observations exist (TDTs 21701113 and 27202302).  However, it is vital
to recognize that corrections for off-source emission in LWS 
are inappropriate for faint objects like normal stars in low-cirrus regions.

All observed LWS spectra have had the instrumental dark currents subtracted from them 
during the reductions.  In our OLP10 reductions, we
used the ``fixed dark currents", as determined in dedicated calibration 
observations\footnote{http://www.iso.vilspa.esa.es/manuals/HANDBOOK/lws\_hb/node30.html}.
These LWS dark current measurements were essentially dark background
measurements. The Fabry-Perot was placed in the beam and offset.  Therefore,
there was flux on the detectors, radiated by the 3.5-K blank, but 
this was negligible. The signals measured remained constant throughout the
mission, and were subtracted from the measured photocurrents on-source. 
When the dark signal, measured from the blank, was compared with the signal measured 
on the dark patch of sky, no difference was found.  Therefore, for dark sky, no 
additional ``sky" subtraction is necessary. For regions with high cirrus,
there is a sky signal in the LWS beam and, if a point source is being
observed, an ``off" position is needed to determine the local sky background. In
dark regions, the sky flux predicted by COBE can form a substantial fraction
of the LWS dark signal.  However, this is deemed undetected, because no additional 
signal is detected over the signal from the blank. 

Consequently, we must assess the measured sky background around the two K~giants at
those times when the LWS spectra were taken.  If those backgrounds were below 
the fixed dark values, then there is no need for further subtraction of off-source
LWS spectra.  Indeed, that subtraction would produce meaningless stellar spectra
by removing the dark currents twice. Both LWS and ISOPHOT, off-source, measure the 
sum of zodiacal (hereafter ``zodi"), cirrus, and extragalactic emission, as did DIRBE.
``IRSKY"\footnote{http://www.ipac.caltech.edu/ipac/services/irsky/irsky.html}
shows that extragalactic and cirrus confusion noises are far below zodi, so
the dominant background for the two K~giants is zodi.
We have used two methods to evaluate the FIR background near $\alpha$~Tau and
$\alpha$~Boo: actual measurements in off-source pixels when these stars were 
observed in the FIR arrays of ISOPHOT; and, as a coarse check, DIRBE predictions 
for the LWS observations based on their solar elongation angles.

The computation of ISOPHOT background sky measurements near $\alpha$~Tau and $\alpha$~Boo 
is summarized in Tables~\ref{ataupht} and ~\ref{aboopht}.  These present the relevant 
ISOPHOT observation 
by filter, nominal wavelength, TDT and measurement number,
and the background brightness and its 1-$\sigma$ uncertainty.  Our conversion from 
point source flux to surface brightness is different from the one given in 
the ISOPHOT handbook \citep{isohandbk}.  The supporting analysis and justification are given
by Schulz\footnote{http://www.iso.vilspa.esa.es/users/expl\_lib/PHT\_list.html/fpsf\_report.ps.gz}.
When multiple data exist for a filter, inverse-variance weighted combinations 
of those data are also given.  Conversion from MJy~sr$^{-1}$ to
F$_\lambda$ (in W~cm$^{-2}$~$\mu$m$^{-2}$) was accomplished using the published
values for LWS detector beam solid angles, effective apertures, and bandwidths.
Table~\ref{bgnd} presents these backgrounds, and compares them with the fixed dark currents 
and DIRBE predictions (interpolated between 140 and 240~$\mu$m to assess the 160-$\mu$m values),
matching each ISOPHOT band to the nearest LWS detector.  The DIRBE predictions are generally 
in accord with ISOPHOT measurements.  Clearly, the measured sky backgrounds around the 
two stars are all below the fixed dark currents, confirming that only low level cirrus 
appears in their vicinities (with the sole exception of $\alpha$ Boo in LW2, where 
ISOPHOT suggests 50\% more sky brightness than the LW2 dark current).  The DIRBE 
predictions are generally in accord with the ISOPHOT measurements.  \emph{Subtraction of 
off-source LWS spectra would be inappropriate for these K~giants}.



\begin{figure}
\plotone{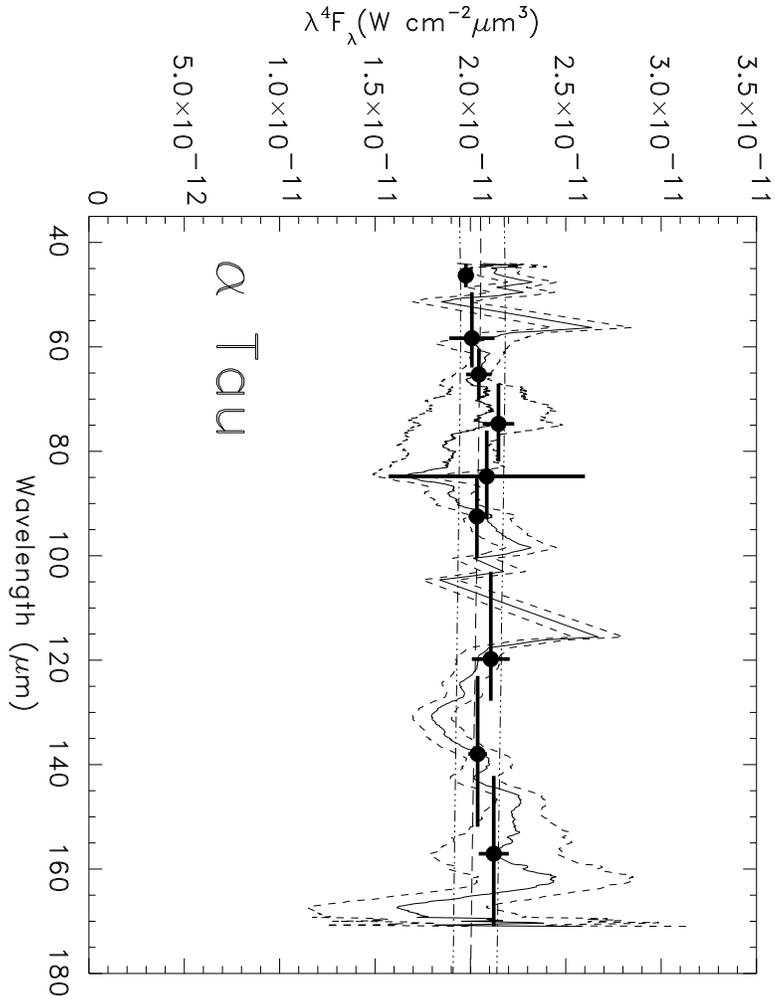}
\caption {LWS spectrum of $\alpha$~Tau showing the mean observed spectrum (solid line)
after boxcar smoothing,
and the mean$\pm3\sigma$ bounds (short-dashed lines), compared with the Plez model continuum
spectrum (long-dashed lines) and with the mean$\pm1\sigma$ bounds on the old
SED provided to ISOPHOT in 1996 (dash-triple-dotted lines). 
Each detector result is shown as a heavy cross, where
horizontal bars represent the actual wavelength extent of the useful data, and
vertical bars show the 1-$\sigma$ errors on the plotted weighted-mean points.}
\label{ataulws}
\end{figure}
\begin{figure}
\plotone{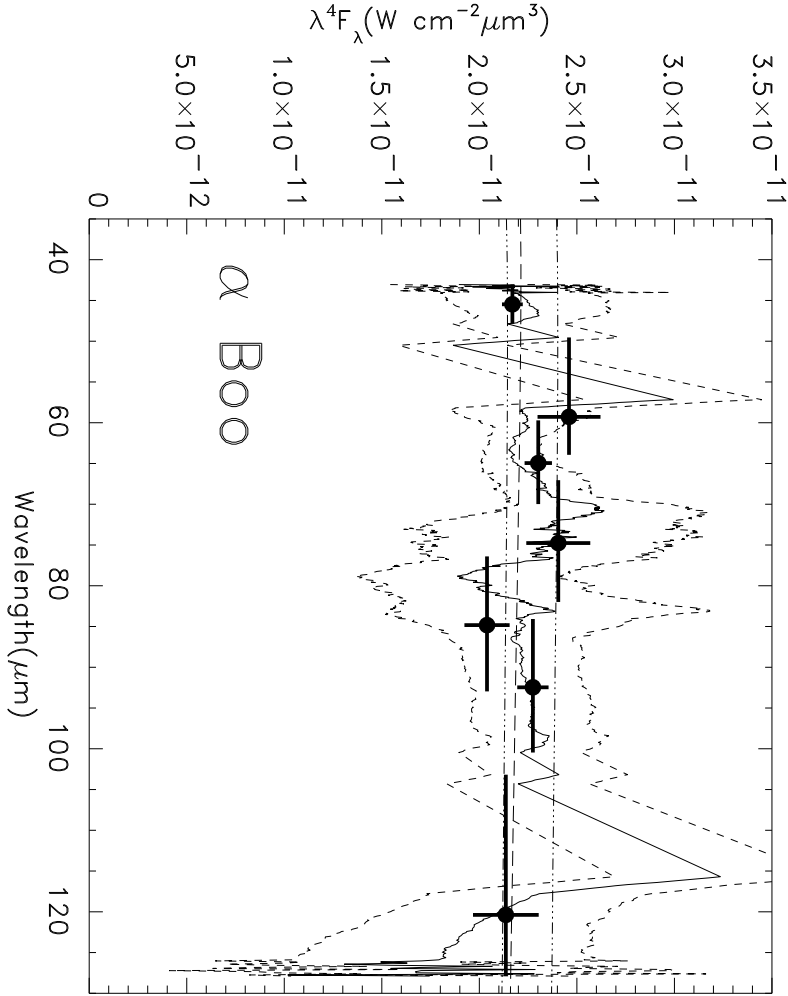}
\caption {LWS spectrum of $\alpha$~Boo showing the mean observed spectrum (solid line)
after boxcar smoothing,
and the mean$\pm2\sigma$ bounds (short-dashed lines), compared with the PDK model continuum
spectrum (long-dashed lines) and with the mean$\pm1\sigma$ bounds on the old
SED provided to ISOPHOT in 1996 (dash-triple-dotted lines). Heavy crosses as for
Fig.~\ref{ataulws}.}
\label{aboolws}
\end{figure}

Figures~\ref{ataulws} and~\ref{aboolws} illustrate our reduced, spliced LWS
spectrum for each star, presenting the mean spectrum over all TDTs, with the $\pm3\sigma$ 
bounds for $\alpha$~Tau (which has more LWS data, longer integrations, and higher SNR) 
and $\pm2\sigma$ bounds for $\alpha$~Boo (with its noisier data sets).  The sizeable
triangular excursions in both spectra near 53 and 106~$\mu$m represent the
difficulty, for NIR-bright normal stars, of mitigating the
substantial leaks of 1.6-$\mu$m radiation to which LWS is prone.
We have cleaned the observed spectra by 30-point boxcar smoothing of 
the data (equivalent to about a 2-$\mu$m interval). $\alpha$~Tau's combined spectrum
contained 1181 wavelength points; $\alpha$~Boo's 706 points.  Each figure incorporates
the absolutely calibrated model continuum spectra that we have computed (see below)
as the almost flat long-dashed lines that cross the figures, and the mean$\pm1\sigma$
bounds on our older calibrated energy distributions (dash-triple-dotted lines)
delivered in January 1996 for the calibration of ISOPHOT.  
To convert the model surface fluxes to predicted fluxes at the Earth, we have
adopted the angular diameter, 20.88$\pm$0.10 mas, of \citet{lunocc}
(using lunar occultation) for $\alpha$~Tau and the angular diameter,
21.0$\pm$0.2 mas, of \citet{quirren} for
$\alpha$~Boo (measured by optical intensity interferometry).

Our new, more detailed, calculations of stellar continuum spectra
fall well within the uncertainties associated with the calibrated spectra
originally delivered to ISO.   We wish to quantify the degree of accord
between measured and modeled spectra in the FIR for each star.
The grating simultaneously illuminated the ten LWS detectors, and each
detector was separately wavelength and flux calibrated.  Therefore, LWS
consists of ten spectrally independent detectors.  The degree of
oversampling of these stellar spectra was 4; consequently, adjacent
points in Figures~\ref{ataulws} and~\ref{aboolws} are not independent.
Therefore, to assess the consistency of observations and predictions,
and to handle the dependence of adjacent LWS points in the fully-plotted
spectra correctly, we include the data binned by detector as a series of bold 
crosses in Figures~\ref{ataulws} and~\ref{aboolws}.  Filled circles represent
the inverse-variance weighted average of all the meaningful data in each LWS 
detector (e.g. after the rejection of data contaminated by
$H$-band leaks).  Error bars show the associated 1-$\sigma$ uncertainties.  
Asymmetric horizontal error bars represent the actual wavelength ranges usable for 
each detector and star.  Filled circles are plotted at the average 
wavelength of each detector's set of usable data for that star.

The binned detector data lie within 1$\sigma$ of the newly modeled energy
distribution for $\alpha$~Tau, and within 1$\sigma$ for $\alpha$~Boo.  There
is good agreement between the underpinning Uranus calibration of LWS, the totally
independently calibrated stellar spectra made from the empirical spectra by our older 
model extrapolations based on the available observations between 8 and 23~$\mu$m,
and the newly calculated, absolutely
calibrated, emergent stellar energy distributions.
Averaging the results for the independent detectors for each
star with inverse-variance weighting, we determine that, 
for $\alpha$~Tau, this mean scale factor is 1.000$\pm$0.011 (over nine
detectors) while, for $\alpha$~Boo, this factor is 1.012$\pm$0.025 (over
seven detectors).  Therefore, empirical stellar and planetary calibrations
used by ISO, and stellar modeling, are absolutely reconciled
to 3.3\% (3$\sigma$) for the better measured $\alpha$~Tau, and 7.5\%
(3$\sigma$) for the more poorly measured $\alpha$~Boo.\\

\section{The Radiative Equilibrium Model Atmospheres }
\subsection{Atmospheric parameters and computation of the emergent spectrum}

The spectrum calculations were carried out with the SOURCE model
atmosphere program in essentially the same fashion as described in
\citet{dfcetal82}.  Specifics pertinent to the current calculations
are described in this section.  

For $\alpha$~Tau, we adopted an effective temperature of 3920~K
\citep{blgp91} and a log(g) of 1.5 \citep{sl90}.  For the isotopic
abundances of C, N and O, we adopted log~$\epsilon$(C)~=~8.40,
log~$\epsilon$(N)~=~8.20, log~$\epsilon$(O)~=~8.78,
$^{12}$C/$^{13}$C~=~10, $^{16}$O/$^{17}$O~=~562, and
$^{16}$O/$^{18}$O~=~475 \citep{hl84,sl90}.  The remaining metal
abundances were chosen to be solar, following the lead of
\citet{sl85}.  The preceding abundances were derived using the MOOG
spectrum synthesis code \citep{sneden73} and model atmospheres
\citep{johnson80} constructed with the ATLAS (e.g., \citet{kurucz93})
model atmosphere code and opacity sampling.  Our choices of T$_{eff}$,
log(g), and abundances are entirely consistent with current
determinations of the parameters of $\alpha$~Tau.  In particular, the
values are well within the error bars of the values recently derived
directly by Decin and her collaborators \citep{decin00,decin03} from
fits to stellar spectral energy distributions generated by using
the MARCS \citep{gustafsson75,plez92} model atmosphere code.  Bertrand
Plez \citep{plez_pc} kindly computed an $\alpha$~Tau model atmosphere
for us using the aforementioned parameters and the programs described
in \citet{plez92}.  Thus, this model was computed with essentially the
same model atmosphere program and opacities as employed by Decin.

For $\alpha$~Boo, we adopted a slightly revised version (\citet{ruth}:
hereafter PDK) of the \citet{peterson} model atmosphere which has more
depths than the original published model.  This atmosphere has an
effective temperature of 4300~K, log(g) of 1.5, [Fe/H]~=~-0.5, and
other elemental abundances as determined by \citet{peterson}.  The PDK
model parameters derived by \citet{peterson} were derived entirely
from the observed stellar spectrum.  As was the case for the Decin
et al. investigation of $\alpha$~Tau, no assumptions were made
regarding stellar temperature or gravity.  We adopted all of the
\citet{peterson} values for our study.  The parameters T$_{eff}$,
gravity, microturbulent velocity, and [Fe/H] derived by
\citet{peterson} for $\alpha$~Boo are essentially identical to those
derived later and independently by \citet{decin03}, while the C, N,
and O abundances agree within the error bounds.

The alert reader will observe that we have adopted model atmospheres
for $\alpha$~Tau and $\alpha$~Boo that were derived using entirely
different model atmosphere codes (ATLAS and MARCS).  We shortly will
describe results obtained by using these atmospheres as input to a
third model atmosphere code, our SOURCE.  We do not feel that significant
improvement would result from insisting that \emph{only} ATLAS
\emph{or} MARCS models be used in our study.  All three codes have
been vetted against the solar case and should give very comparable
results in the G-K star domain.  We cite as support for this stance
the striking agreement between the results of \citet{peterson} and
\citet{decin03} for $\alpha$~Boo.  This occurs despite \citet{decin03}
having used a completely different model atmosphere program and
blanketing representation (MARCS with opacity sampling) to compute
their $\alpha$~Boo model than used by \citet{peterson} (ATLAS with
opacity distribution functions).  The agreement is even more
impressive when one considers that these two investigations derived
their results from two entirely different spectral regions, 0.5--0.9~$\mu$m \emph{versus} 2.38--12.0~$\mu$m.  We regard this as
evidence that different well-vetted model atmosphere codes and opacity
representations now produce comparable model atmospheres and
visible-infrared spectra for K-giants \textbf{\emph{within the limited
approximations}} of static, homogeneous atmospheres, LTE, and
mixing-length-theory radiative/convective equilibrium.  That does
\textbf{\emph{not}} mean that the resultant models are accurate, or
even adequate, representations of a real star.  Differences between
observed and computed spectra are now much more likely to arise from
the presence of physics (\emph{e.g.}, NLTE, non-radiative heat
deposition, and inhomogeneous dynamical structure) \textbf{\emph{not treated}}
in the simple models we use here.  We will discuss this point in more
detail as we proceed.

The column mass (g~cm$^{-2}$) {\em vs.\/} temperature relations for
both the Plez $\alpha$~Tau and PDK $\alpha$ Boo models were linearly
extrapolated to shallower depths as required to obtain optical
transparency at the longest wavelengths.  For the Plez $\alpha$~Tau
model, the extrapolated layers are shallower than 0.020 g~cm$^{-2}$;
for the PDK $\alpha$~Boo model, shallower than 0.010 g~cm$^{-2}$. The
resultant model atmospheres are shown in
Figure~\ref{atau_aboo_models}.

It should be stressed that both of the above adopted model atmospheres
have purely radiative equilibrium surface temperature structures.
Specifically, they do {\em not\/} have chromospheric temperature
rises, a point whose importance will become evident later.  We have
included in Figure~\ref{atau_aboo_models}, for comparison,
the \citet{kelch} (KEL) $\alpha$~Tau chromospheric model B and the
\citet{al} (AL) $\alpha$~Boo chromospheric model.  These chromospheric
models are semiempirical, derived from non-LTE analyses of \ion{Ca}{2}
and \ion{Mg}{2} line fluxes.  While these chromospheric models are now
over 20 years old, they are very similar from the upper photosphere,
through the temperature minimum region, and into the lower
chromosphere to models of $\alpha$~Tau and $\alpha$~Boo used by
\citet{wied}, and the $\alpha$~Tau model derived by \citet{mcmurry}.

Using the Plez $\alpha$~Tau atmosphere and parameters detailed above,
we have computed the synthetic spectrum for $\alpha$~Tau over the
entire LWS range and beyond, 43 to 665~$\mu$m.  The spectrum was
computed using a full set of molecular line opacities which includes
all the principal isotopomers of CO, SiO, OH, and ${\rm H_{2} O}$ and
using a variable wavelength mesh which locally provides $\approx$2.5
points per Doppler width.  To allow comparison with the LWS
observations, we convolved the high-resolution result with a Gaussian
instrumental profile function having a FWHM equal to 0.29~$\mu$m for
$\lambda \leq$ 90.5~$\mu$m, and 0.60~$\mu$m, $\lambda >$ 90.5~$\mu$m
(e.g., \citet{gry}).  The resulting spectrum, normalized to the
continuum and on a greatly magnified ordinate scale, is shown in
Figure~\ref{atau_spectrum}.  A detailed description of the displayed
spectrum may be found in \citet{carbon05}.  We do note here that
almost all the absorption beyond $\approx$~50~$\mu$m is due to the
isotopomers of SiO.  At shorter wavelengths the principal contributor
is OH with some H$_{2}$O and CO as well.

It is clear that, {\em at the ISO LWS and the BIMA spectral
resolutions\/}, all differences between the pure continuum and the
synthetic spectrum computed with the full set of molecular line
opacities are less than 2\%; beyond 130~$\mu$m, they are less than
1\%.  This is {\em much less\/} than the noise level in the observed
$\alpha$~Tau spectra in this region.  As a consequence, there is no
significant advantage gained by using the full synthetic spectrum.
For clarity in our presentation, we will use only the continuous
spectrum in our comparisons with the $\alpha$~Tau observations.  Since
$\alpha$~Boo is both hotter and has lower metallicity, we shall use
the continuum alone for comparisons there as well.

\subsection{Depth of formation and brightness temperatures}

\begin{figure}
\plotone{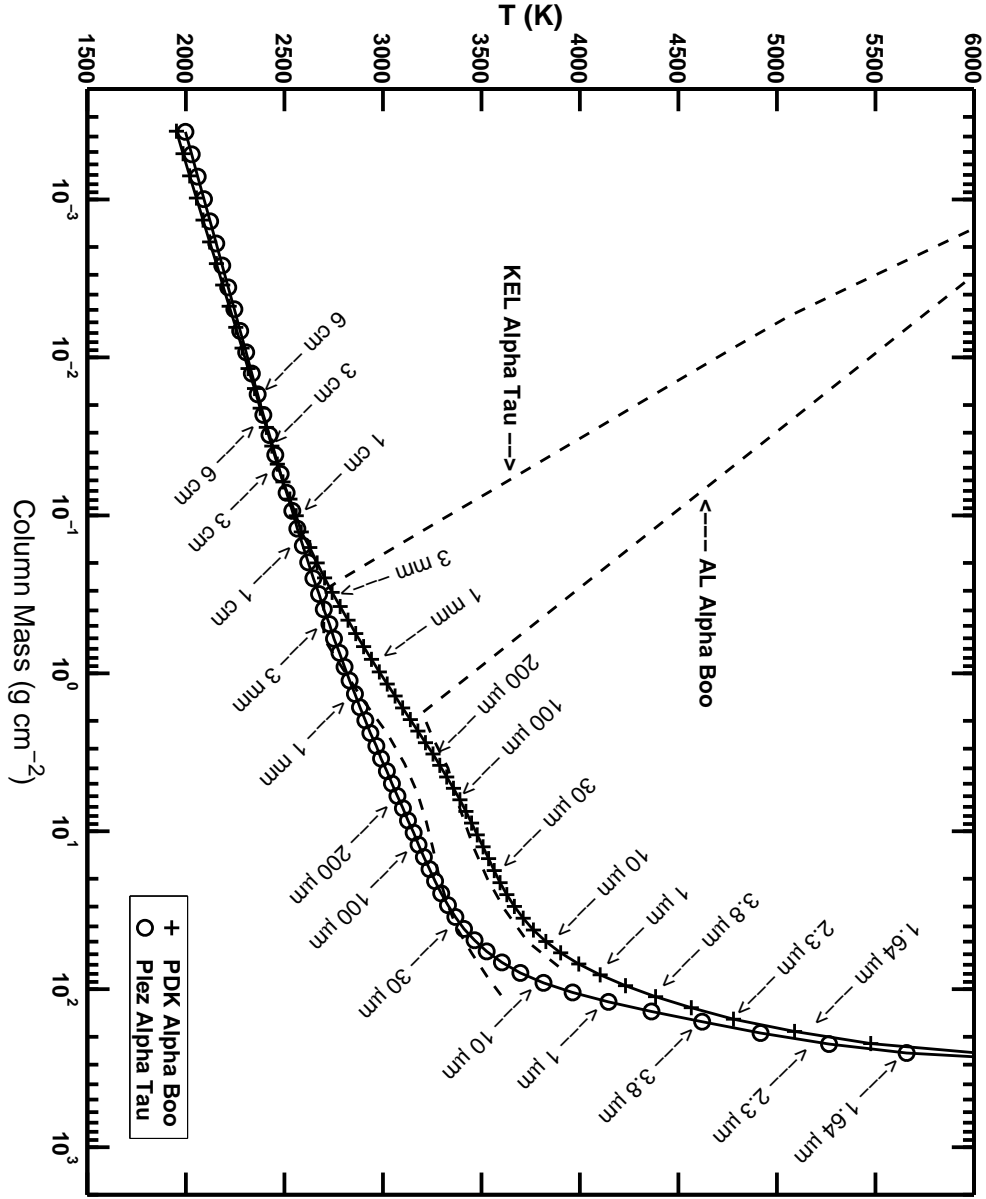}
\caption {The adopted model atmosphere temperature structures for
$\alpha$~Boo and $\alpha$~Tau.  Arrows mark the depths at which
optical depth unity is reached in the continuum at the indicated
wavelengths.  The upper set of arrows refers to the PDK model; the lower set, to the 
Plez model.  For comparison, the AL and KEL chromosphere models are shown as 
dashed lines. }
\label{atau_aboo_models}
\end{figure}

\begin{figure}
\plotone{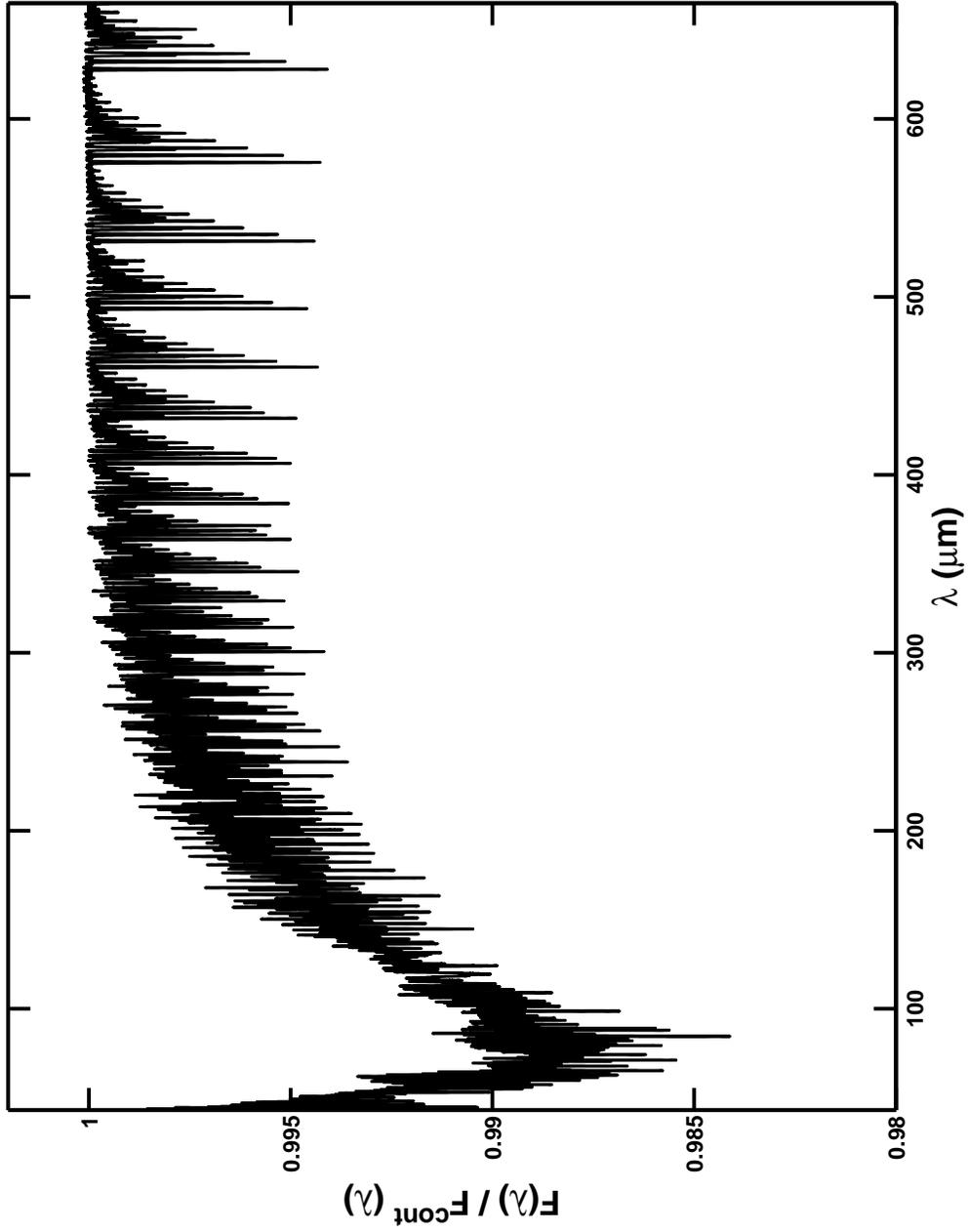}
\caption {Continuum normalized flux predicted by the Plez $\alpha$~Tau  model for the spectral range 
43~$\mu$m to 665~$\mu$m.  Note that the vertical scale is greatly magnified.}
\label{atau_spectrum}
\end{figure}

For both the Plez $\alpha$~Tau and the PDK $\alpha$~Boo models, the
dominant continuous opacity in the continuum-forming layers for
wavelengths redward of the H$^-$ opacity minimum at 1.64~$\mu$m arises
from H$^-$ free-free.  This opacity increases in strength as
$\lambda^2$.  As a result, the emergent continuum flux is formed at
progressively shallower layers in the stellar atmosphere as one
proceeds to longer wavelengths.  Figure~\ref{atau_aboo_models}
illustrates this point for the Plez $\alpha$~Tau and the PDK
$\alpha$~Boo models.  In this figure we have marked, for the radiative
equilibrium models, the atmospheric depths for which the continuum
optical depth, $\tau_{\lambda}^{c}$, reaches unity at the indicated
wavelengths.  That is, we have marked the column masses at which
$\tau_{\lambda}^{c}$~=~1.0; roughly speaking, in the FIR,
depths near this value and shallower make the greatest contribution to
the emergent continuum flux at wavelength $\lambda$.  Note, for
example, that the continuum fluxes between 1~mm and 3~mm are sampling
very different parts of the atmospheric temperature structure in both
radiative equilibrium models than, say, the continuum fluxes at
2.3~$\mu$m.  Indeed, the longest wavelengths are sampling, {\em in the
continuum\/}, atmospheric levels reached only at the cores of lines in
bluer portions of the spectrum.  Only when one proceeds to the vacuum
ultraviolet and its strong photoionization continua of metals is it
possible to reach in the continuum the shallow atmospheric levels
probed by the continuum at long wavelengths (e.g. Figure 1 of
\citet{val76}).  Thus, the FIR continuum is an effective probe of the
shallowest stellar layers which can supplement information derived
from line cores and UV continua.

As noted above, there is very little difference between the full
synthetic spectrum and the spectrum of the continuum.
Figure~\ref{atau_aboo_models} provides an important clue as to why
spectral lines generally become less prominent at longer wavelengths.
It is evident that, as one goes to shallower depths in the models, the
slope of the temperature-column mass relation becomes smaller.  The
significance of this can be appreciated by considering an absorption
line in the $\alpha$~Tau model with a characteristic,
depth-independent, line core/continuum opacity ratio of 1000.  If the
absorption line occurs at a wavelength of 2.3~$\mu$m, the continuum
will be formed at $\approx$5200~K and the line core will be formed at
$\approx$3200~K.  The same line at a wavelength of 200~$\mu$m will
have the continuum formed at $\approx$3100~K and the line core at
$\approx$2600~K, a much smaller temperature difference between line
core and continuum than at 2.3~$\mu$m.  Couple this with the
decreasing temperature sensitivity of the Planck function as one goes
to longer wavelengths and one has a recipe for progressively less
prominent spectral features.

\begin{figure}
\plotone{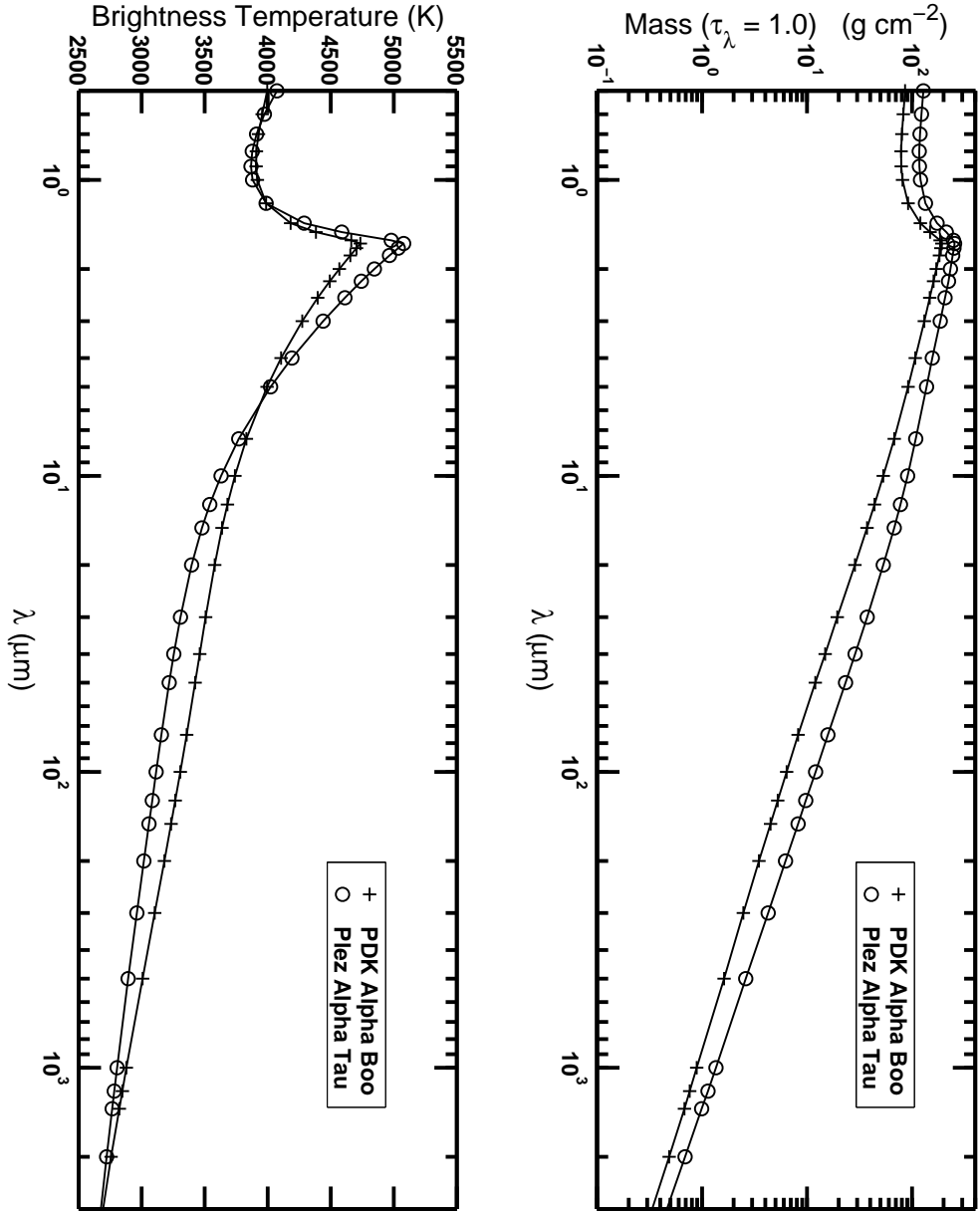}
\caption {Upper panel: The column masses at which continuum
optical depth unity is reached in the PDK $\alpha$~Boo and Plez $\alpha$~Tau
radiative equilibrium models.  Lower panel: The brightness temperatures predicted by the
PDK $\alpha$~Boo and Plez $\alpha$~Tau radiative equilibrium models.}
\label{mass_BT}
\end{figure}

Figure~\ref{mass_BT} provides another window into the wavelength
dependence of continuum formation.  In the upper panel of
Figure~\ref{mass_BT} we show the column masses for which
$\tau_{\lambda}^{c}$~=~1.0 over the spectral range 0.5~$\mu$m
to 200~$\mu$m.  As noted above, there is a steady march to
progressively shallower depths of continuum formation as one goes to
longer wavelengths.  Since the dominant H$^-$ bound-free opacity peaks
at roughly 0.825~$\mu$m, continuum fluxes at wavelengths between 0.825~$\mu$m 
and 1.64~$\mu$m probe the same atmospheric layers as continuum
fluxes between 1.64~$\mu$m and 6.5~$\mu$m.  Longward of 6.5~$\mu$m,
the continuum flux probes atmospheric layers accessible only in line
cores and in the metal bound-free continua of the vacuum ultraviolet.

In the lower panel of Figure~\ref{mass_BT} we show the continuum
``brightness temperature" over the same spectral range for both
models.  The brightness temperature is defined as the temperature of a
blackbody which gives the same flux as the model atmosphere at the
indicated wavelength (a precise definition is given in
Section~\ref{allbt}). Three points are important.  First, neither
$\alpha$~Tau nor $\alpha$~Boo could have its continuum fluxes over any
large spectral range matched by a blackbody of a single temperature.
Second, the blackbody temperatures that do match the model fluxes at
specific wavelengths are generally quite different than the effective
temperatures of the models (or the stars) themselves.  Third,
comparing with Figure~\ref{atau_aboo_models}, one sees that the
brightness temperature is systematically smaller than
T($\tau_{\lambda}^{c}$~=~1.0), the temperature of the atmospheric
layer at which optical depth unity is reached in the continuum at
wavelength $\lambda$. (For an atmosphere with a linear source function at wavelength $\lambda$,
the brightness temperature at $\lambda$~and~T($\tau_{\lambda}^{c}$~=~2/3) would be
identical.  Unfortunately, in the particular cases of the PDK and Plez models, the FIR 
source functions are not linear in $\tau_{\lambda}^{c}$ but, rather, quadratic.)  The
differences between T($\tau_{\lambda}^{c}$~=~1.0) and brightness
temperature vary systematically with wavelength.  The differences are
greatest at 1.64~$\mu$m and decrease to the red.  \emph{Beyond
40~$\mu$m, the differences are always less than 100~K.\/} (Had
we chosen for illustration the more correct, but \emph{still} approximate,
T($\tau_{\lambda}^{c}$~=~2/3), the temperature differences would be
smaller yet.)  Thus, the brightness temperature is a good first
guess at T($\tau_{\lambda}^{c}$~$\approx$~1.0) at the longer
wavelengths. Later, when we discuss differences between observed and
predicted brightness temperatures, the reader may use this
approximation and Figure~\ref{atau_aboo_models} to roughly estimate the implied errors in the radiative equilibrium model
temperature structures.

\section{Discussion}

\subsection{Evidence for significant flux excesses at 1.4~mm and 2.8~mm}\label{excess}

Table~\ref{compare} shows the comparison between the observed fluxes
and the predictions of the radiative equilibrium models.
Only for the case of $\alpha$~Tau at 1.4~mm does there appear to be
approximate agreement between the observed flux and the flux predicted
from the radiative equilibrium model.  The remainder of the observed
fluxes are all significantly {\em larger\/} than the fluxes predicted by
the models.  The discrepancies are particularly striking in the case
of $\alpha$~Boo.


We do not believe that the flux computation for the radiative
equilibrium models could have errors sufficiently large to produce the
marked differences between computed and observed millimeter fluxes.
Our model atmosphere code has been carefully vetted and the dominant
continuous opacities are themselves quite well determined (e.g.,
\citet{firacc}).  (As an example of evidence for this, we refer the
reader forward in the text to Figure~11.  There we show
the continuum for the Plez $\alpha$~Tau  model computed with an entirely
independent model code.)  On the observational side, our derived
errors for the observed 1.4-mm and 2.8-mm fluxes preclude dismissing
the differences between observed and computed fluxes as simply due to
measurement error.  We believe other explanations must be considered.

Since the adopted angular diameters for these two stars appear to be
extremely well determined, with standard deviations of $\approx$1\%
or less, they would not seem to be a significant source of error.
Moreover, a major change in the angular diameters would compromise
flux agreement for the $\lambda \leq$ 200~$\mu$m region.
Nevertheless, it should be noted that the \citet{quirren} diameter is
an arithmetic mean of values measured from 0.45~$\mu$m to 2.2~$\mu$m and
the \citet{lunocc} diameter is a error-weighted mean biased toward
values obtained at 1.6~$\mu$m and 3.8~$\mu$m.  At any particular
monochromatic wavelength, an observed angular diameter represents the
apparent diameter of the $\tau_{\lambda}$~$\approx$~1.0 surface of the
star.  Since, as shown in Figures~\ref{atau_aboo_models} and~\ref{mass_BT}, 
there is a progressive march in depth of continuum
formation to ever shallower atmospheric layers as one moves redward of
1.64~$\mu$m, the angular diameters themselves should increase with
wavelength.  If the angular diameters at 1.4~mm and 2.8~mm are
substantially larger than those at the wavelengths of the
\citet{quirren} and \citet{lunocc} observations, our predicted fluxes
in Table~\ref{compare} could be seriously underestimated.  We can
roughly estimate the magnitude of this source of error, at least in
the case of the radiative equilibrium models.  In Table~\ref{extents}
we show the physical radii of $\alpha$~Tau and $\alpha$~Boo based on
the indicated angular diameters and Hipparcos parallaxes
\citep{hip}.  We also show the physical distances in the radiative
equilibrium models between $\tau_{1.64 \mu m}^{c}$~=~1.0 and
$\tau_{2.8 mm}^{c}$~=~1.0.  


Given the tabulated numbers, it is clear that, for the PDK and Plez
models, the atmospheric extent in the continuum at 2.8~mm is not
sufficient to seriously affect our predicted fluxes.  We find a
similar result in the case of the only model chromosphere with a
published height scale, McMurry's chromospheric model for $\alpha$~Tau
\citep{mcmurry}.  The McMurry atmosphere is plane-parallel and
hydrostatic, constructed with log g = 1.25.  Note that this gravity
leads to a more distended atmosphere than does the higher gravity we
adopted.  The McMurry atmosphere has an extent of
$\approx$2.6~$\times$~10$^{11}$~cm between the layers where the
1.64~$\mu$m continuum is formed and the layer with T~=~7120~K in the
chromosphere.  This extent is only 8\% of the stellar radius and again
seems too small to explain the discrepancy between computed and
observed fluxes.  Finally, as we will describe in Section~\ref{allbt},
\citet{drake} determined that $\alpha$~Boo has essentially the photospheric
radius at 2~cm.  Given the expected increase in angular
diameter with wavelength, this would imply that $\alpha$ Boo would
have the photospheric radius at $<$~2~cm as well.

Based on the above arguments, we believe that the flux excesses
observed at 1.4~mm and 2.8~mm, relative to the predictions of the
purely radiative models, are real and significant.  Furthermore, as we
shall show, there is good evidence that these excesses are signatures of
the chromospheric temperature rise in the two red giants.

\subsection{Other observations of the sub-mm through cm region}
\label{mmcm}
\begin{figure}
\plotone{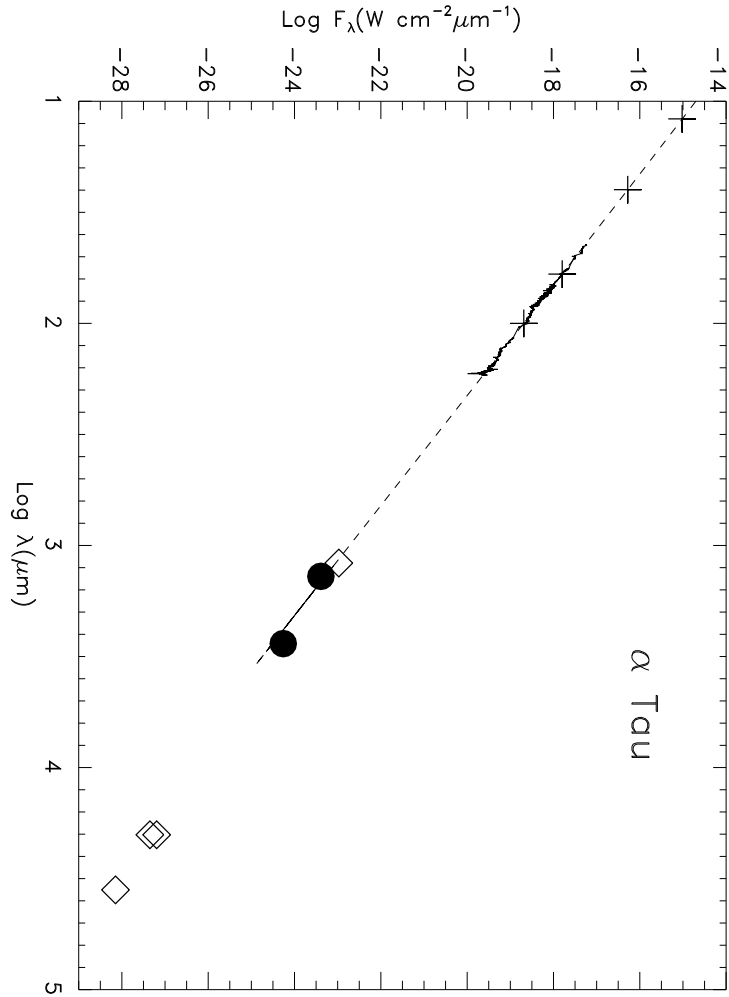}
\caption {FIR to cm spectrum of $\alpha$~Tau.  Color-corrected 
IRAS points (crosses) are plotted together with our mm observations (filled 
circles), the LWS spectrum (solid), and our computed continuum spectrum 
(dashed line).  Also shown are mm and cm data from the literature 
(open diamonds).}
\label{ataummcm}
\end{figure}

\begin{figure}
\plotone{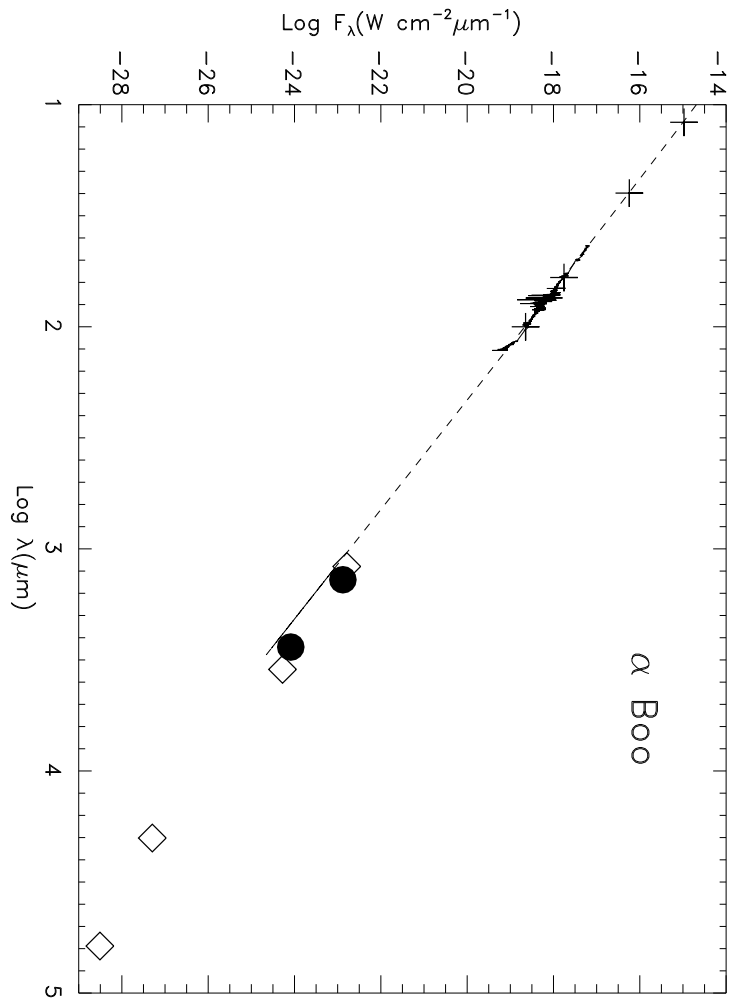}
\caption {FIR to mm spectrum of $\alpha$~Boo as for Figure~\ref{ataummcm}.}
\label{aboommcm}
\end{figure}

Figures~\ref{ataummcm} and~\ref{aboommcm} present our millimeter flux
densities (filled circles) and the color-corrected IRAS measurements
of each star at 12, 25, 60, and 100~$\mu$m (crosses).  Each figure
includes the Plez $\alpha$~Tau or PDK $\alpha$~Boo spectrum,
as appropriate, from 10~$\mu$m to 3~mm.

There are additional high-frequency radio continuum observations of these
two K giants in the literature.  \citet{altcomet} made single-dish 86-GHz
measurements of $\alpha$~Boo (21.4$\pm$7.5 mJy) with the IRAM 30-m telescope,
while \citet{alt250} likewise detected both $\alpha$~Tau and $\alpha$~Boo
at 250 GHz, obtaining 51$\pm$6 and 78$\pm$8 mJy, respectively.  There are
also VLA centimeter measurements on both stars at 4.9, 8.4, and 14.9 GHz by
\citet{drake}, and unpublished data by Drake, Linsky, \& Judge from
the compilation of radio data by \citet{wendk}.  These long wavelength
measurements all have substantial uncertainties, in part due to
the faintness of the stars, in part to the potential for confusion with
extragalactic background objects.  Nonetheless, in combination
with our own radio data (Figures~\ref{ataummcm},~\ref{aboommcm}) one sees
that these high-frequency data (open diamonds) are in good agreement with our 
aperture synthesis measurements (filled circles).  

\subsection{Brightness temperatures for all the observed fluxes}\label{allbt}

\begin{figure}
\plotone{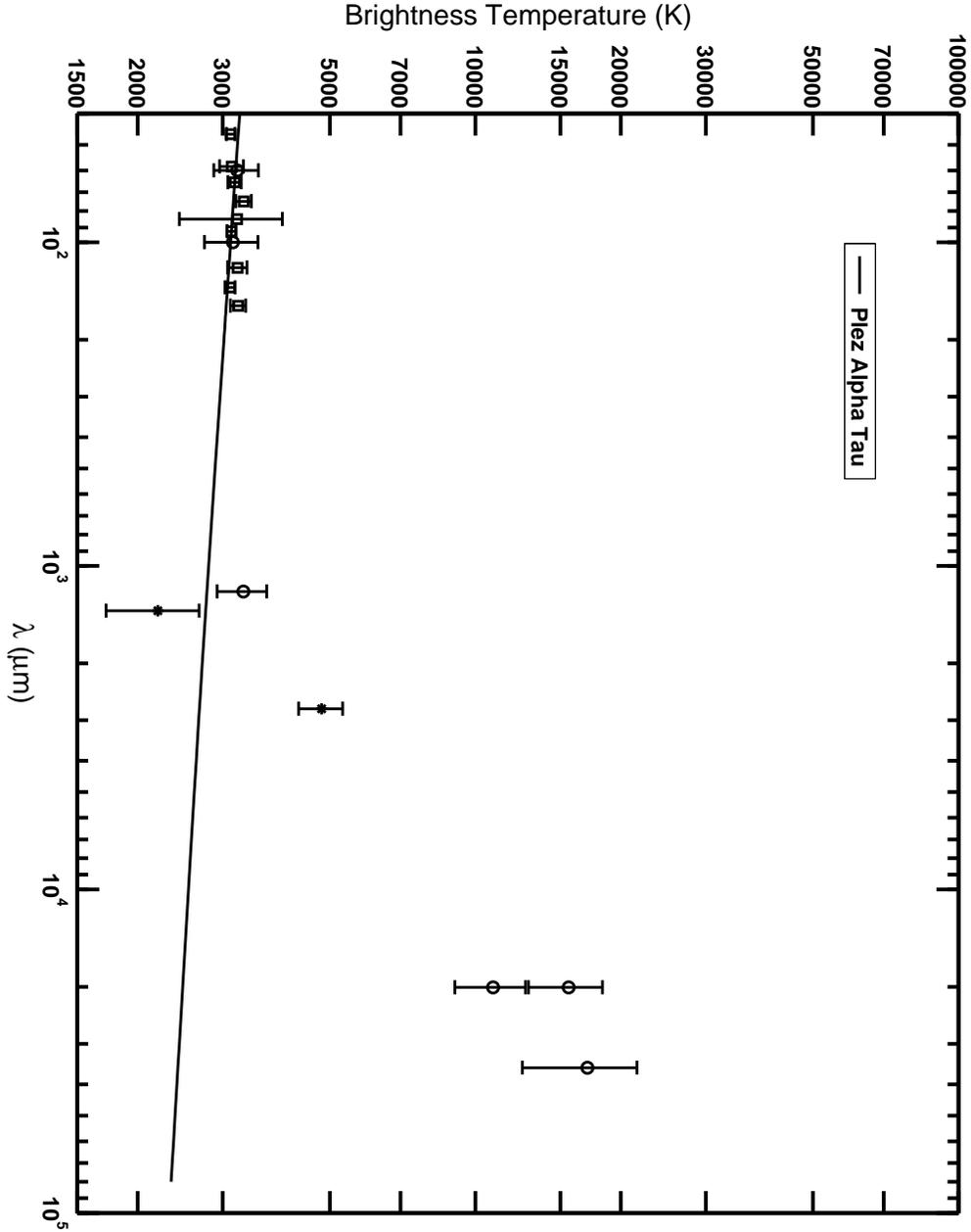}
\caption {Brightness temperatures corresponding to the observed and computed fluxes for $\alpha$~Tau 
which have been discussed in this paper.  LWS observations are represented by open 
squares, our mm measurements by asterisks, all other observations by open 
circles. Error bars represent 1-$\sigma$ deviations in the fluxes only; errors for 
2-cm and 3.6-cm points are our own estimates (see text). Brightness temperatures 
predicted by the Plez model are shown by a solid line. Note that the brightness temperatures at cm wavelengths may be only upper limits.  See the text for a discussion of the impact of wavelength dependent angular diameters.}
\label{atauallbt}
\end{figure}

\begin{figure}
\plotone{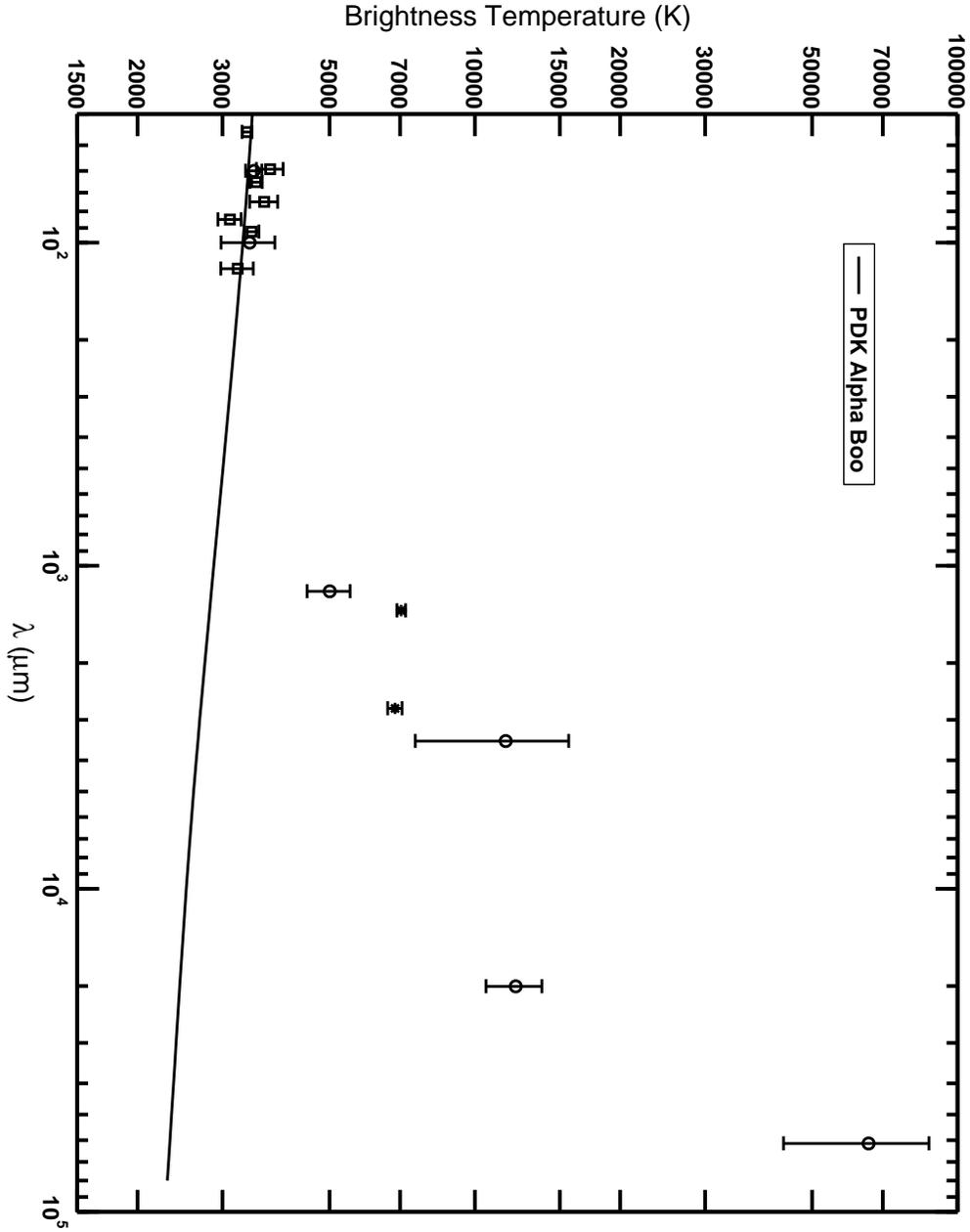}
\caption {Brightness temperatures corresponding to the observed and computed fluxes for $\alpha$~Boo 
which have been discussed in this paper.  LWS observations are represented by open 
squares, our mm measurements by asterisks, all other observations by open 
circles. Error bars represent 1-$\sigma$ deviations in the fluxes only. Brightness 
temperatures predicted by the PDK model are shown by a solid line.  Note that the brightness temperatures at cm wavelengths may be only upper limits.  See the text for a discussion of the impact of wavelength dependent angular diameters.}
\label{abooallbt}
\end{figure}

It is difficult to fully appreciate the significance of the excess
fluxes from the logarithmic scales of Figures~\ref{ataummcm}
and~\ref{aboommcm}.  An insightful alternative view may be obtained
by combining the mm and cm fluxes from other observers with our data
and looking at the whole dataset as brightness temperatures over a
broad spectral range.  In Figures~\ref{atauallbt} and~\ref{abooallbt}
we show the brightness temperatures corresponding to all the observed
flux data between 43~$\mu$m and 6~cm used in our previous figures.
For comparison we have also plotted the predictions of the radiative
equilibrium models.  In deriving brightness temperatures from the
observed fluxes, we used the angular diameters of Table~\ref{extents}
in the following relation:
\begin{displaymath}
T_{B}(\lambda) = \frac{14387.75\:/\:\lambda}{\ln(\:1 \;+\; \frac{733.4090 \: \alpha^{2}}{S_{Jy}(\lambda) \: \lambda^{3}})}
\end{displaymath}
where $S_{Jy}(\lambda)$ is the observed flux in Jy, $\alpha$ is the
angular diameter in mas, $\lambda$ is the wavelength in microns, and
$T_{B}(\lambda)$ is in Kelvins.  The error bars
in Figures~\ref{atauallbt} and~\ref{abooallbt} correspond \emph{only\/} to the
uncertainties in the observed fluxes.  In the case of the $\alpha$~Tau
data at 2~cm and 3.6~cm, no error values were reported by
\citet{wendk} and we used our own rough estimates of the plausible
errors.
 Before proceeding, a caveat needs to be noted regarding the longer
 wavelength points especially.  The observed fluxes presumably come
 from progressively shallower stellar layers as $\lambda$ increases.
 All of the brightness temperatures in Figures~\ref{atauallbt}
 and~\ref{abooallbt} have been deduced assuming the photospheric
 angular diameters of Table~\ref{extents}.  Atmospheric extension may
 play a role in the chromospheric layers of $\alpha$~Tau and
 $\alpha$~Boo.  \citet{carpenter} estimated the extension of the
 chromospheres of these stars. They argued that the \ion{C}{2} UV 0.01
 lines which they studied were formed in chromospheric layers with
 temperatures of $\approx$10000~K in $\alpha$~Tau and $\approx$8000~K
 in $\alpha$~Boo.  They deduced that these chromospheric layers had
 radii of $\approx$2.0~R$_{\ast}$ and $\approx$1.4~R$_{\ast}$,
 respectively, where R$_{\ast}$ is the stellar photospheric radius.
 Although these estimates are, in the authors' words, ``relatively
 crude'', they do suggest that atmospheric extension may play a role
 at least in the hotter layers of the chromosphere.  Different
 evidence of extension was found by \citet{drake} in their study of
 2~cm and 6~cm fluxes.  They treated all the radio
 emission as coming from an optically thick ionized wind for which a
 ``half-power radius'' could be deduced.  In the case of $\alpha$~Boo
 they found the 2~cm half-power radius was $\approx$R$_{\ast}$; at
 6~cm, the half-power radius had grown to $\approx$1.7~R$_{\ast}$.
 Taking the long wavelength approximation to the brightness
 temperature expression above, we see that, for a given observed flux,
 $T_{B}(\lambda) \propto~1 / \alpha^{2}$, where $\alpha$ is the
 angular diameter.  Since the shallowest layers may have angular
 diameters significantly larger than those of the photosphere, the
 corresponding brightness temperatures in Figures~\ref{atauallbt}
 and~\ref{abooallbt} should be regarded as upper limits.

\subsection{Evidence for contributions from a stellar chromosphere}\label{chrom}
We now describe why we believe that the flux excesses at the longer
wavelengths arise from the presence of stellar chromospheres in $\alpha$~Boo and $\alpha$~Tau.

\subsubsection{Circumstantial Evidence}\label{chrom_cir}

Circumstantial evidence suggests the possibility that stellar
chromospheres could explain the flux excesses.  This evidence comes by
examining the depths at which $\tau_{\lambda}^{c}$~=~1.0.  In the case
of $\alpha$~Boo, as shown in Figure~\ref{atau_aboo_models}, the AL
chromospheric model departs from the purely radiative equilibrium
model at a column mass of $\approx$~0.75~g~cm$^{-2}$ \citep{al}.  This
is just in the column mass range of the PDK radiatve equilibrium model
where optical depth unity is reached between 200~$\mu$m and 1~mm.  Given
this, it is not surprising that the PDK model might significantly
\emph{underpredict} the flux at both 1.4~mm and 2.8~mm in
$\alpha$~Boo.  We believe that the progressively larger differences
between the predictions of the PDK model and the observations beyond
120~$\mu$m are due to the contributions of $\alpha$~Boo's chromosphere.

The circumstantial evidence in the case of $\alpha$~Tau  is not so clear.  As
shown in Figure~\ref{atau_aboo_models}, the KEL $\alpha$~Tau
chromospheric model begins its temperature rise at a column mass of
$\approx~$0.3~g~cm$^{-2}$ \citep{kelch}.  This is slightly
\emph{shallower} in the Plez $\alpha$~Tau model than where
$\tau_{3~mm}^{c}$~=~1.0 occurs.  Nevertheless, the 2.8~mm observed
brightness temperature is distinctly higher than that predicted by the
Plez model in Figure~\ref{atauallbt}.  In fact, with the exception of
our 1.4~mm point, the general run of observed points beyond 200~$\mu$m
suggests that some enhanced (chromospheric?) emission may be present
at 1~mm or even shorter wavelengths.  Whether this is true or not will
take additional observations in this spectral region with better SNR.

\subsubsection{Evidence from NLTE modeling}\label{chrom_nlte}

\begin{figure}
\plotone{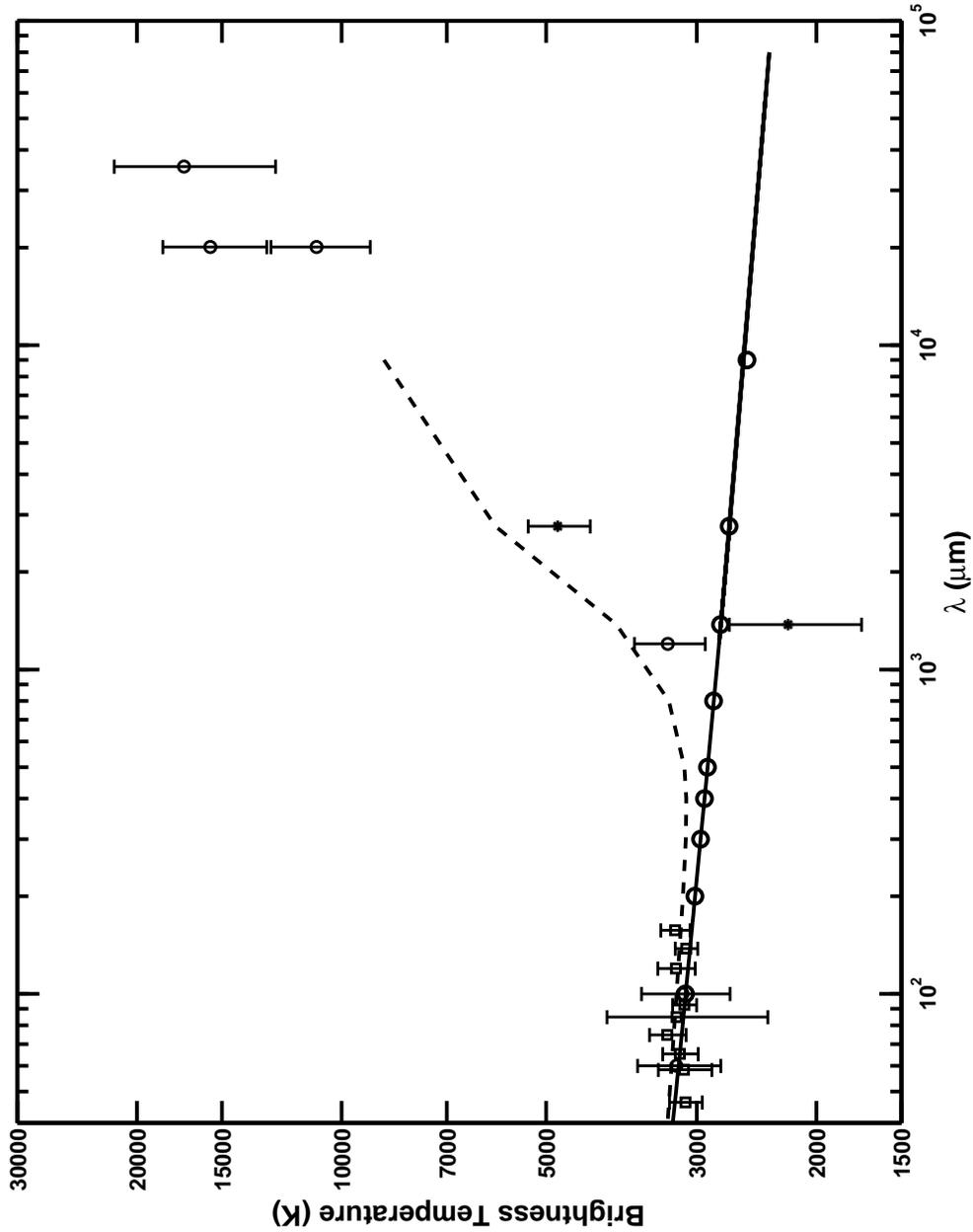}
\caption {Brightness temperatures for $\alpha$~Tau including the predictions from the McMurry NLTE 
chromospheric model and from an independent computation using the LTE Plez model. Observations are 
marked exactly as in Figure~\ref{atauallbt} with the same error bars.  We have magnified the vertical 
scale slightly to reduce the clutter of points. The brightness temperatures predicted by the McMurry 
NLTE chromospheric model are indicated as a dashed line.  Brightness temperatures for the Plez model 
in LTE computed using MULTI are shown as open circles \emph{without} error bars.  Brightness temperatures 
predicted by SOURCE from the Plez model are shown by a solid line as before. 
Note that the brightness temperatures at cm wavelengths may be only upper limits.  
See the text for a discussion of the impact of wavelength dependent angular diameters. }
\label{atau_nlte}
\end{figure}

While NLTE chromospheric modeling is well beyond the intent of this
investigation, one of us (ADM) has computed the infrared fluxes
predicted by his chromospheric model of $\alpha$~Tau .  Details of the model
and the NLTE computation may be found in \cite{mcmurry}.  This model
is essentially \emph{identical} to the KEL model from the photosphere
through the lower and middle chromosphere, differing from the KEL
model only at column masses shallower than~10$^{-3}$~g~cm$^{-2}$ (see
Fig.~5, \cite{mcmurry}).  The NLTE radiative transfer computation was
carried out with the independent code MULTI.  We show in
Figure~\ref{atau_nlte} the results of this NLTE computation with a
chromospheric model compared with the observations, and with the
predictions of the Plez model.

It is immediately apparent that, again with the exception of our
1.4~mm point, the observed data beyond 200~$\mu$m is better fit by the
NLTE chromosphere model than the LTE Plez radiative equilibrium model.
We believe this is good evidence that the stellar chromosphere of
$\alpha$~Tau has a significant impact on the observed fluxes beyond 200~$\mu$m.
Indeed, it should be noted that the fit is even improved between 
100~$\mu$m and 200~$\mu$m.

\subsection{The transition from photosphere to chromosphere}
\label{photchrom}

For $\alpha$~Boo (Figure~\ref{abooallbt}) it appears that the transition from
purely radiative occurs beyond 120~$\mu$m, and certainly well before
1.4~mm.  In our opinion this makes any purely radiative model for
$\alpha$~Boo  a poor choice for calibration in the sub-mm spectral region.
This appears to be the case for $\alpha$~Tau as well, especially considering
the predictions of the McMurry chromospheric model
(Figure~\ref{atau_nlte}).  Given the evidence described above, we do
not believe that it is prudent to use cool star radiative models for
calibration purposes at the longer wavelengths \emph{unless
alternatively calibrated observations already confirm} the validity of
the purely radiative models.  One must be certain that the flux
predicted by the purely radiative model at the long wavelength end of
the region being calibrated by the theoretical stellar spectrum does
indeed correspond to the actual star.  In other words, we believe that
when one is approaching the spectral regions where a chromospheric
contribution can be significant, one can safely use purely radiative
cool star models only as an interpolation tool for calibrations, not as
an extrapolation tool.

We would like to emphasize the importance of the spectral region from
$\approx$30~$\mu$m to several millimeters to furthering an
understanding of cool stellar atmospheres.  Fluxes from this spectral
region reflect the mean time- and spatial-averaged emission from the
upper photosphere and chromosphere for stars like those we consider
here (see Figure~\ref{atau_aboo_models}).  Accurate fluxes measured
through this spectral region should allow one to test directly model
atmosphere structures, even complex ones that include NLTE and
dynamics, over the very interesting depth range in which non-radiative
processes like shocks begin to alter the structure.  By selectively
sampling progressively longer wavelengths, one can follow the important
transition from photosphere to chromosphere, locating the depth at
which purely radiative models like the PDK and Plez models begin to
fail.




\subsection{Stellar absolute calibrators for ISOPHOT}
Our original, theoretically extrapolated, continuum spectra for the two
K giants, supplied to ISOPHOT as absolutely calibrated energy distributions,
are in extremely good agreement with our new calibrated synthetic spectra
(Figures~\ref{ataulws},~\ref{aboolws}), well within the $\pm$6\% uncertainties
assigned to the original calibrators \citep{vii}.
The LWS spectra, reduced with respect to Uranus as calibrator, are
consistent with the new (and old) synthetic spectra,
thereby uniting FIR planetary and stellar calibrations
at the few percent level (Figures~\ref{ataulws},~\ref{aboolws}).

There are
independent data on $\alpha$~Boo, derived from the validation of ISOPHOT's final 
pipeline (OLP10) \citep{phil}, that show it conforms to our 1996 model spectrum.
In addition, \citet{phttab} has described the performance through the ISO mission
of a significant number of ISOPHOT calibrators.  By removing the contributions
made by any particular source to the calibration process, and calibrating that
source with respect to all other calibrators, one can assess the ratio of
observed to modeled irradiance.  Using the database described by \citet{phttab},
we have been able to extend the range
over which our new model of $\alpha$~Boo can be tested, and to validate $\alpha$
Tau independently of our LWS data.  For $\alpha$~Boo, we have determined the
observed to modeled irradiance ratios for the ISOPHOT C100, C120, C135, C160, C180, and C200
filters (``C" denotes the FIR camera within the ISOPHOT instrument, and
each filter is designated by its reference wavelength in~$\mu$m).  Expressing
these observed irradiance ratios with respect to our new modeled spectrum for $\alpha$~Boo
we find:  C100, 1.02$\pm$0.02; C120, 1.14$\pm$0.08; C135, 1.01$\pm$0.02; 
C160, 1.06$\pm$0.02; C180, 1.00$\pm$0.03; C200, 1.00$\pm$0.26.  For $\alpha$~Tau, 
measurements are available solely in the C160 filter and yield a ratio of 
0.96$\pm$0.05, with respect to our new model.  We conclude that the separate 
absolute calibrations of LWS and ISOPHOT are in accord within the uncertainty of the 
observations, despite their different dependences on planets, asteroids, and stars.  
This should not be interpreted to mean that more accurate observations might not reveal 
discrepancies between FIR observations and the predictions of static, LTE model atmospheres 
computed in simple radiative/convective equilibrium.

\section{Conclusions}

Our primary objective was to investigate whether purely radiative
models for two normal K giants could reproduce the observed FIR and
microwave fluxes.  We have found that for both $\alpha$~Tau and
$\alpha$~Boo the radiative equilibrium temperature structures of the
upper photospheres produce fluxes in agreement with the observations,
but only up to a certain point.  For $\alpha$~Boo and $\alpha$~Tau the
agreement extends to at least 125~$\mu$m, corresponding to a column
mass of $\approx$5~gm~cm$^{-2}$ in the PDK atmosphere and
$\approx$10~gm~cm$^{-2}$ in the Plez atmosphere.  Both K giants
clearly show definite chromospheric emission at 2.8~mm.  In 
$\alpha$~Boo this is evident by 1.4~mm.  Only additional observations
can pinpoint precisely where between 125~$\mu$m and 2.8~mm the
chromospheric contribution becomes significant.  For this reason, we
cannot rely on purely radiative model atmospheres to form a basis for
stellar calibration in the sub-mm region.  Purely
radiative models of the continuum might be used to provide
calibrations \emph{between} those shorter FIR wavelengths where the validity 
of the model fluxes already has been independently verified by observations.


Finally, our results indicate that accurate FIR, sub-mm, and mm
observations can provide valuable new information for researchers
modeling stellar chromospheres of cool stars.

\acknowledgments
We wish to thank Bertrand Plez for computing an $\alpha$~Tau model
for our particular set of input parameters and Ruth Peterson for the
extended version of the \citet{peterson} $\alpha$~Boo model.  Their help is 
greatly appreciated.  It is a pleasure to thank Ullie Klaas and 
Phil Richards of the ISOPHOT team for sending us data on their validation of OLP10 
as it bears on $\alpha$~Boo's spectrum, and Sumita Jayaraman for assistance
in extracting DIRBE data for the elongation angles at the times of LWS and
ISOPHOT observations of these K~giants.
MC acknowledges support for this work through subcontract UF99025 between the 
University of Florida and UC Berkeley, supporting the calibration of ISOPHOT.
DFC gratefully acknowledges access to the resources of the NASA
Advanced Supercomputing (NAS) Division, NASA/Ames Research Center,
Moffett Field, CA, as well as the support of the NAS Division Chief
Walter Brooks and NAS Branch Chiefs Dochan Kwak and William
W. Thigpen.  SGI/Cray C90 supercomputer cycles for this research were
provided through the Consolidated Supercomputing Management Office
(CoSMO) at NASA/Ames.  This material is based upon work by DFC and DG supported
by the National Aeronautics and Space Administration under NRA-99-01-ADP-073
issued through the Astrophysics Data Program.  We acknowledge support for
BIMA and CARMA through NSF grants AST-9981308 and AST-0228963.  



\epsscale{0.8}

\clearpage
\oddsidemargin=-1cm
\tabletypesize{\scriptsize}

\begin{deluxetable}{cccccccccc}
\tablewidth{0pt} 
\tablenum{1}
\tablecolumns{9}
\tablecaption{Journal of observations\label{obs}}
\tablehead{Star&  Date&  Array&  Frequency& Time&   Phase& Secondary &   Primary&  Traceable& Peak F$_{\nu}$\\   
               &      &       &        GHz& on star& Ref.&  Flux Ref.& Flux Ref.& to Primary& mJy\\}
\startdata
$\alpha$ Tau& 14Nov97& C& 106.6845& 4.32 hr& 0449+113& 3C111& Mars, & \nodata& 17.14$\pm$2.19\\
\nodata&  \nodata& \nodata&   \nodata& \nodata&  \nodata&  \nodata& Uranus, Neptune& \nodata& \nodata\\
$\alpha$ Boo& 15Nov97& C& 106.6845& 4.22 hr& 1415+133& 3C111& Mars, & \nodata& 20.27$\pm$2.95\\
\nodata&  \nodata& \nodata& \nodata& \nodata& 1357+193& \nodata& Uranus, Neptune& \nodata& \nodata\\
$\alpha$ Boo& 15Nov97& C& 110.1282& 4.22hr& 1415+133& 3C111& Mars, & \nodata& 18.47$\pm$2.73\\
\nodata&  \nodata& \nodata& \nodata& \nodata& 1357+193& \nodata& Uranus, Neptune& \nodata& \nodata\\
$\alpha$ Boo& 15Nov97& C& Both& \nodata& \nodata&  \nodata&  \nodata& \nodata& 19.73$\pm$2.16\\
$\alpha$ Tau& 06Jun98& C& 106.6845& 2.99hr& 0449+113& 3C111& Mars& \nodata& 13.10$\pm$3.40\\
$\alpha$ Boo& 11Mar99& B& 106.6845& \nodata& 1357+193& 1415+133& \nodata& Mars& 24.57$\pm$1.20\\
$\alpha$ Boo& 11Mar99& B& 110.1282& 4.37 hr& 1357+193& 1415+133& \nodata& Mars& 19.68$\pm$1.23\\
$\alpha$ Boo& 11Mar99& B& Both& \nodata& \nodata&  \nodata&  \nodata& \nodata& 22.03$\pm$0.97\\
$\alpha$ Boo& 12Mar99& B& 106.6845& 3.13 hr& 1357+193& 1415+133& MWC~349, Mars& Mars& 21.02$\pm$1.74\\
$\alpha$ Boo& 12Mar99& B& 110.1282& \nodata& 1357+193& 1415+133& MWC~349, Mars& Mars& 18.90$\pm$1.20\\
$\alpha$ Boo& 12Mar99& B& Both& \nodata& \nodata& \nodata&  \nodata& \nodata& 19.92$\pm$1.02\\
$\alpha$ Tau& 05Sep99& D& 216.0982& 1.85 hr& 0530+135& 0530+135& \nodata& MWC~349/Mars& 24.23$\pm$6.87\\
$\alpha$ Tau& 05Sep99& D& 219.5418& \nodata& 0530+135& 0530+135& \nodata& MWC~349/Mars& 27.34$\pm$8.12\\
$\alpha$ Tau& 05Sep99& D& Both& \nodata& \nodata& \nodata&  \nodata& \nodata& 25.78$\pm$5.64\\
$\alpha$ Boo& 01Jun00& D& 216.0982& 2.91 hr& 3C273& 3C273& MWC~349& Mars& 81.55$\pm$2.35\\
$\alpha$ Boo& 01Jun00& D& 219.5418& 2.91 hr& 3C273& 3C273& MWC~349& Mars& 85.45$\pm$2.44\\
$\alpha$ Boo& 01Jun00& D& Both& \nodata& \nodata& \nodata&  \nodata& \nodata& 83.50$\pm$1.71\\
$\alpha$ Tau& 09Dec00& C& 106.6845& 3.19 hr& 3C111& 3C111& W3OH, Mars,& \nodata& 12.55$\pm$2.16\\
\nodata&  \nodata& \nodata&   \nodata& \nodata&  \nodata&  \nodata& Uranus& \nodata& \nodata\\
$\alpha$ Tau& 09Dec00& C& 110.1282& 3.19 hr& 3C111& 3C111& W3OH, Mars,& \nodata& 12.77$\pm$2.95\\
\nodata&  \nodata& \nodata&   \nodata& \nodata&  \nodata&  \nodata& Uranus& \nodata& \nodata\\
$\alpha$ Tau& 09Dec00& C& Both& \nodata& \nodata& \nodata&  \nodata& \nodata& 12.40$\pm$1.91\\
\tableline
$\alpha$ Tau& All& All& 108.40& \nodata& \nodata& \nodata&  \nodata& \nodata& 13.97$\pm$1.46\\
$\alpha$ Tau& All& All& 217.82& \nodata& \nodata& \nodata&  \nodata& \nodata& 25.78$\pm$5.64\\   
$\alpha$ Boo& All& All& 108.40& \nodata& \nodata& \nodata&  \nodata& \nodata& 20.09$\pm$0.69\\
$\alpha$ Boo& All& All& 217.82& \nodata& \nodata& \nodata&  \nodata& \nodata& 83.50$\pm$1.71\\
\enddata
\end{deluxetable}

\begin{deluxetable}{lcccc}
\tablewidth{0pt} 
\tablenum{2}
\tablecolumns{5}
\tablecaption{ISOPHOT measurements of sky background near $\alpha$~Tau \label{ataupht}}
\tablehead{Filter& $\lambda$& TDT& Meas.\#& Background$\pm$Unc.\\
 & ($\mu$m)& & & (MJy~sr$^{-1}$)\\}
\startdata                                                  
C1-100& 100& 86401601& 2& 42.32$\pm$3.57\\
C1-100& 100& 86401602& 2& 44.53$\pm$2.68\\
C2-160& 170& 86002102& 1& 43.06$\pm$4.39\\
C1-60\tablenotemark{a}& 60& 84300501& 5& 34.54$\pm$4.68\\
C1-100\tablenotemark{a}& 100& 84300501& 2& 36.69$\pm$4.14\\
C1-60\tablenotemark{b}& 60& 84300501& 5& 40.89$\pm$5.53\\
C1-60& 60& combined& \nodata& 37.19$\pm$3.57\\
C1-100& 100& combined& \nodata& 42.24$\pm$1.90\\  
\enddata
\tablenotetext{a}{P32}
\tablenotetext{b}{scaled P32} 
\end{deluxetable}
 
\begin{deluxetable}{lcccc}
\tablewidth{0pt}
\tablenum{3}
\tablecolumns{5}
\tablecaption{ISOPHOT measurements of sky background near $\alpha$~Boo \label{aboopht}}
\tablehead{Filter& $\lambda$& TDT& Meas.\#& Background$\pm$Unc.\\
 & ($\mu$m)& & & (MJy~sr$^{-1}$)\\}
\startdata                                                  
C1-50& 50& 27501511& 1& 18.16$\pm$1.42\\
C1-90& 90& 27502117& 1& 14.62$\pm$0.87\\
C1-105\tablenotemark{a}& 105& 27500602& 1& 12.54$\pm$0.54\\
C1-105& 105& 27500602& 1& 11.56$\pm$0.94\\
C2-120& 120& 27503008& 1& 9.10$\pm$1.16\\
C2-135& 135& 27503311& 1& 8.73$\pm$1.12\\
C2-160& 170& 27503614& 1& 9.23$\pm$0.96\\
C2-160\tablenotemark{a}& 170& 27503614& 1& 8.34$\pm$0.27\\
C2-200& 200& 27502402& 1& 6.61$\pm$0.39\\
C1-105& 105& combined& \nodata& 12.30$\pm$0.47\\  
C2-160& 170& combined& \nodata& 8.41$\pm$0.26\\
\enddata
\tablenotetext{a}{Central pixel only} 
\end{deluxetable}

\begin{deluxetable}{lcllll}
\tablewidth{0pt} 
\tablenum{4}
\tablecolumns{6}
\tablecaption{Comparison of ISOPHOT sky backgrounds and fixed dark currents in W~cm$^{-2}$~$\mu$m$^{-1}$\label{bgnd}}
\tablehead{Star&  ISOPHOT&  Nearest&  Fixed& ISOPHOT& DIRBE\\   
          &   Band&   LWS det.& Dark& Sky&   zodi\\}
\startdata
$\alpha$~Tau& C1-60&  SW2& 4.2E-18&  4.5E-19&  3.8E-19\\
$\alpha$~Tau& C1-100& LW1& 8.0E-19&  1.3E-19&  1.1E-19\\
$\alpha$~Tau& C2-160& LW4& 9.1E-20&  4.5E-20&  5.9E-20\\
\tableline
$\alpha$~Boo& C1-50&  SW2& 4.2E-18&  2.0E-19&  $>$1.3E-19\\
$\alpha$~Boo& C1-90&  SW5& 3.2E-18&  8.7E-20&  \nodata\\
$\alpha$~Boo& C1-105& LW1& 8.0E-19&  5.2E-20&  2.7E-20\\
$\alpha$~Boo& C2-120& LW2& 7.7E-21&  1.2E-20&  1.2E-20\\
$\alpha$~Boo& C2-135& LW3& 2.0E-20&  1.7E-20&  7.0E-21\\
$\alpha$~Boo& C2-160& LW4& 9.1E-20&  1.5E-20&  1.8E-20\\
$\alpha$~Boo& C2-200& LW5& 4.4E-19&  8.6E-21&  \nodata\\
\enddata
\end{deluxetable}
 
\begin{deluxetable}{llcc}
\tablewidth{0pt} 
\tablenum{5}
\tablecolumns{3}
\tablecaption{ Radiative models: fluxes and brightness temperatures \label{compare}}
\tablehead{Frequency& Quantity& \multicolumn{1}{c}{$\alpha$ Boo}& \multicolumn{1}{c}{$\alpha$ Tau}\\}
\startdata                                                  
217.820 GHz (1.376 mm)& Observed Flux (mJy)& 83.50$\pm$1.71&     25.78$\pm$5.64\\
\nodata& Predicted Flux (mJy)&   33.43&               32.38\\
\nodata& $\mid$Predicted--Observed Flux$\mid$/$\sigma$(observed)& 29.3&  1.17\\
& & & \\
\nodata& Observed Brightness Temperature (K)& 7040$\pm$140& 2200$\pm$480\\
\nodata& Predicted Brightness Temperature (K)& 2820& 2760\\
\tableline
108.406 GHz (2.765 mm)& Observed Flux (mJy)&      20.09$\pm$0.69&      13.97$\pm$1.46\\
\nodata& Predicted Flux (mJy)&       7.94&                7.80\\
\nodata& $\mid$Predicted--Observed flux$\mid$/$\sigma$(observed)& 17.6&  4.23\\
& & & \\
\nodata& Observed Brightness Temperature (K)& 6840$\pm$230& 4810$\pm$500\\
\nodata& Predicted Brightness Temperature (K)& 2700& 2690\\
\enddata
\end{deluxetable}

\begin{deluxetable}{llcc}
\tablewidth{0pt} 
\tablenum{6}
\tablecolumns{3}
\tablecaption{Radiative Models: atmospheric extents \label{extents}}
\tablehead{Quantity& \multicolumn{1}{c}{$\alpha$ Boo}& \multicolumn{1}{c}{$\alpha$ Tau}\\}
\startdata                                                  
Adopted angular diameter (mas)&                   21.0$\pm$0.2&        20.88$\pm$0.10\\
Adopted parallax (mas)&                           88.85$\pm$0.74&      50.09$\pm$0.95\\
Deduced stellar radius (cm)&                      1.77 x 10$^{12}$&      3.12 x 10$^{12}$\\
1.64 $\mu$m - 2.8 mm atmospheric thickness (cm)&  4.38 x 10$^{10}$&      4.22 x 10$^{10}$\\
Percentage of stellar radius&                     2.5\%&              1.4\%\\
\enddata
\end{deluxetable}
\end{document}